\documentclass[10pt,twocolumn,notitlepage,sort&compress]{article}
\usepackage [english]{babel}
\usepackage[utf8]{inputenc}

\usepackage{graphicx} 
\usepackage{dcolumn} 
\usepackage{bm} 
\usepackage{amssymb}
\usepackage{amsmath}
\usepackage{subcaption}
\usepackage{xcolor}
\usepackage[margin=0.5in]{geometry}
\usepackage{caption}
\usepackage{setspace}
\usepackage{multicol}
\usepackage{authblk}
\usepackage{xfrac}
\usepackage{multirow,tabularx}
\usepackage{soul}

\usepackage[resetlabels,labeled]{multibib}
\newcites{Supp}{References}

\PassOptionsToPackage{hyphens}{url}
\usepackage{hyperref}

\hypersetup{
    colorlinks,
    citecolor=blue,
    filecolor=red,
    linkcolor=blue,
    urlcolor=blue
}

\newcommand{\paren}[1]{\left( #1 \right)}
\newcommand{\sqp}[1]{\left[ #1 \right]}

\newcommand{\absv}[1]{\left| #1 \right|}

\newcommand{\North}{\mathrm{N}}
\newcommand{\South}{\mathrm{S}}

\newcommand{\ddx}[1]{\frac{\mathrm{d}}{\mathrm{d} #1}}
\newcommand{\ppx}[1]{\frac{\partial}{\partial #1}}

\newcommand{\be}{\begin{equation}}
\newcommand{\ee}{\end{equation}} 
\newcommand{\lb}{\label}
\newcommand{\OL}{\overline}
\newcommand{\cO}{{\mathcal O}}

\newcommand{\bu}{{\bf u}}
\newcommand{\grad}{{\mbox{\boldmath $\nabla$}}}

\newcommand{\bdot}{{\mbox{\boldmath $\cdot$}}}

\begin{document}



\title{Global Cascade of Kinetic Energy in the Ocean and the Atmospheric Imprint}

\author[1]{Benjamin A. Storer}
\author[2]{Michele Buzzicotti}
\author[3]{Hemant Khatri}
\author[4]{Stephen M. Griffies}
\author[1]{Hussein Aluie\thanks{hussein@rochester.edu}}


\affil[1]{Department of Mechanical Engineering and Laboratory for Laser Energetics, University of Rochester, Rochester, New York, USA}

\affil[2]{Department of Physics, University of Rome Tor Vergata and INFN, Rome, Italy}

\affil[3]{Department of Earth, Ocean, and Ecological Sciences, University of Liverpool, UK}

\affil[4]{NOAA Geophysical Fluid Dynamics Laboratory and Princeton University Atmospheric and Oceanic Sciences Program, Princeton, New Jersey, USA}

\date{\today}

\maketitle

{\bf 
We present the first estimate for the ocean's global scale-transfer of kinetic energy (KE), across scales from 10~km to 40000~km. We show the existence of oceanic KE transfer between gyre-scales and mesoscales induced by the atmosphere's Hadley, Ferrel, and polar cells, and intense downscale KE transfer associated with the Inter-Tropical Convergence Zone. We report peak upscale transfer of 300 GigaWatts across mesoscales of 120~km in size, roughly 1/3rd the energy input by winds into the oceanic general circulation. This ``cascade'' penetrates almost the entire water column, with nearly three quarters of it occurring south of 15$^\circ\South$. The mesoscale cascade has a self-similar seasonal cycle with characteristic lag-time of $\mathbf{\approx27~}$days per octave of length-scales such that transfer across 50~km peaks in spring while transfer across 500~km peaks in summer. KE content of those mesoscales follows the same self-similar cycle but peaks $\mathbf{\approx40~}$days after the peak cascade, suggesting that energy transferred across a scale is primarily deposited at a scale 4$\times$ larger.
}

\section*{\MakeUppercase{Introduction}}
Oceanic general circulation is a central component of Earth's climate system, without which much of the Earth's surface would be covered by ice \cite{winton2003climatic}. This circulation comprises motions spanning a wide range of structures and scales from $\cO(10^4)~$km down to $\cO(1)~$mm, including coherent jets such as the Gulf Stream and Kuroshio, gyres, and the meridional overturning circulation on basin scales several thousands of kilometers in extent \cite{talley2011descriptive}. Ocean circulation also includes turbulent mesoscale eddies of $\cO(100)~$km in size, which pervade the global ocean and contain most of the ocean's kinetic energy \cite{Storer2022}.

Mesoscale eddies are the ocean's equivalent of weather systems, with characteristic timescales of a few months \cite{Chelton2011}. Due to their energy and chaotic nature, recent studies \cite{Arbic2014,serazin2018inverse,arzel2020contributions,gehlen2020quantification2,Juling2021,constantinou2021intrinsic,hochet2022energy} have suggested that these eddies may play a substantial role in climate variability that is intrinsic to the ocean, and is distinct from variability due to external forcing of the ocean. Such oceanic internal variability is hypothesized to arise because energy can be transferred between seemingly incoherent mesoscale eddies and the larger scale coherent flow \cite{Vallis2006}, which evolves on the long timescales of climate and is directly coupled to it. Below, we provide direct evidence of such transfer.

The kinetic energy (KE) cascade, conceptualized by Richardson, Kolmogorov, and Onsager \cite{richardson1922weather,Kolmogorov41a,Onsager49,eyink2006onsager}, is a fundamental  process in turbulent flows with profound and far-reaching consequences, including in our everyday lives. The cascade allows the transfer of energy between vastly different length-scales and is still an active research topic \cite{AlexakisBiferale,zhou2021turbulence}. Since oceanic circulation on scales of $\cO(100)~$km and larger is predominantly geostrophic, similar to 2-dimensional flows, it is theoretically predicted to transfer KE upscale \cite{fjortoft1953changes,Kraichnan1967,charney1971geostrophic}. However, these theories have been formulated for idealized homogeneous turbulence without boundaries. Flow in the ocean is highly inhomogeneous, with prominent roles played by continental boundaries, bottom topography, winds, and a plethora of multiscale processes.

How important is the upscale cascade pathway of KE from the mesoscales of size $\cO(10^2)~$km? How does it compare to other energy sources and sinks in the oceanic circulation? Answering these questions is important to determine the energy cascade's potential contribution to climate variability \cite{serazin2018inverse,Juling2021,constantinou2021intrinsic}. Quantifying the oceanic energy cascade is also pertinent to a longstanding problem in physical oceanography of how mesoscales gain and lose their energy. Our limited understanding of these sources and sinks contributes to large uncertainties in the oceanic KE budget \cite{Ferrari2009}.

While we have global estimates of other processes, such as wind forcing \cite{Wunsch1998,Hughes2008} and dissipation to bottom drag and wave generation \cite{arbic2009estimates,Nikurashin2011}, we do not yet have global estimates for the KE cascade. 
This absence is because the KE cascade is inherently a multi-scale process, which requires decomposing the ocean flow at different length scales in a realistic ocean setting. 
Important progress has been made in this regard, since the seminal work of Scott \& Wang \cite{Scott2005}. 
However, these past investigations of the KE cascade have been limited to analysis of small regions due to their reliance on Fourier transforms in a box \cite{Arbic2013,qiu2014seasonal,sasaki2014impact,khatri2018surface} or a traditional turbulence approach using so-called `structure functions' \cite{Balwada2022}. 
Limitations of these approaches have prevented us from both (i) estimating the global KE cascade rate and (ii) from probing length-scales larger than a few hundred kilometers. 
Storer \textit{et al.} \cite{Storer2022} recently showed that regional analysis misses the gyre-scale components of the oceanic circulation altogether, including a previously unrecognized spectral peak due to the Antarctic Circumpolar Current (ACC). 
Compared to energy spectra, analysis of the energy cascade within boxes suffers from compounded uncertainty due to the elimination of gyre-scales, which introduces uncontrolled errors to the calculation of the cascade even at length-scales smaller than the box size \cite{Aluie2018}. 

In this paper, we use a recent coarse-graining methodology \cite{Aluie2018,Storer2022} that frees us from these limitations while conserving energy throughout the global ocean, which is not possible via the approaches of Fourier boxes or structure functions \cite{Storer2022}. 
At gyre-scales, the KE scale-transfer shows signatures of the global atmospheric circulation patterns, i.e. Hadley, Ferrel, and polar cells, through five latitudinal bands of alternating upscale and downscale energy transfer due to an exchange with the mesoscales. We find that the atmosphere's Inter-Tropical Convergence Zone (ITCZ) produces a band of intense downscale KE transfer in the ocean near the equator.
The method allows us to calculate global KE spectra at different depths, which show that mesoscales $\cO(10^2)~$km penetrate the entire water column whereas the circulation at gyre-scales $>10^3~$km weakens notably with depth. 
We also report the first estimate of $300~$GW for the global upscale cascade of mesoscale KE, which demonstrates that it is a substantial energy pathway in the ocean. We find that the seasonal cycle of the mesoscale KE cascade exhibits a characteristic lag-time of ${\approx27~}$days per octave of length-scales such that, in both hemispheres, the KE transfer across ${50~}$km peaks in spring while transfer across ${500~}$km peaks $\approx3$ months later, in the summer. KE content at these length-scales follows a similar seasonal cycle but with the peak KE occurring $\approx41~$days after the peak KE transfer, suggesting that the seasonal variability in the KE spectrum is caused, at least in part, by the upscale KE cascade.  
In this work, we restrict use of the term `cascade' to where the KE transfer is scale-local, reverting to the more general `scale-transfer' when scale-locality has not been determined. In particular, we use `cascade' when discussing the KE mesoscale transfer and present evidence for its scale-locality.

\section*{\MakeUppercase{Results}}

Our methodology is applied to a \(\sfrac{1}{12}^\circ\) state-of-the-art global ocean reanalysis dataset (hereafter `NEMO') that assimilates data from sea-surface temperature, sea level anomaly, in-situ temperature and salinity profiles, and sea ice concentration.
In contrast to the recent analysis in \cite{Storer2022}, here we use the full velocity, including the geostrophic and ageostrophic components. We also analyze global data from AVISO satellite observations, which are limited to the geostrophic component and are included in the Supplemental Material (SM).
Coarse-graining is performed on a range of length-scales, \(\ell\), spanning \(10~\)km to the equatorial circumference of the Earth (denoted with \(\ell_{\ominus}\sim40\times10^3~\)km). 
Details on coarse-graining, the dataset, and the presented diagnostics can be found in the \hyperref[Methods]{Methods} section.
Our results revolve around two key quantities. 
First is the filtering spectrum \cite{Sadek2018}, \(\OL{E}(k_\ell)=\mathrm{d}\mathrm{KE}^{>\ell}/\mathrm{d}k_{\ell}\),
which measures spectral KE density as a function of length-scale, where \(k_{\ell}=\ell^{-1}\) is filtering wavenumber and \(\mathrm{KE}^{>\ell}\) is the KE contained at all scales larger than \(\ell\).
The second diagnostic is  KE scale-transfer (or cascade), \(\Pi_\ell\), which measures the amount of KE transferred from scales larger than \(\ell\) to scales smaller than \(\ell\), and is signed so that a positive/negative value indicates a downscale/upscale energy transfer.

\subsection*{\MakeUppercase{Surface Kinetic Energy Spectra}}
\begin{figure}[tbhp]
    \centering
    \includegraphics[width=8.66cm]{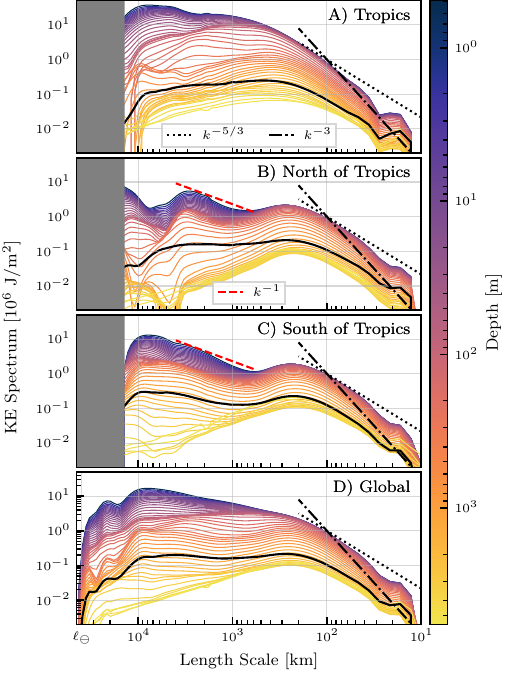}
    \caption{
        \textbf{Kinetic Energy Filtering Spectra:} 
        Area-averaged KE spectra [MegaJoules per square meter; divide by \(1025~\mathrm{kg}/\mathrm{m}^3\) to obtain traditional \(\mathrm{m}^3/\mathrm{s}^2\)] as a function of  length-scale (horizontal axis) and depth (colour scale) for four regions of interest (see panel labels).
        Dashed/dotted black and red lines show various power law slopes, as indicated by the legends.
        Note that the horizontal axis, \(\ell\), decreases to the right. 
        \(\ell_{\ominus}\) denotes the equatorial circumference of the Earth, \(\approx40\times10^3~\)km.
        All 50 vertical levels of the reanalysis dataset are plotted.
        For comparison, solid black curves show KE spectra of the depth-averaged flow.
        See also Figure~\ref{fig:spectra:SUPP} in the SM, which presents spectra at only a few selected depths.
        In [A-C], the largest scales [shaded] cannot be restricted to separate regions.
    }
    \label{fig:surface_spectra}
\end{figure}

Figure~\ref{fig:surface_spectra} presents the area-averaged KE filtering spectra as a function of depth and lateral length-scale for the global ocean (`Global'), north of \(15^\circ\North\) (`North of Tropics', NH), between \(15^\circ\South\) and \(15^\circ\North\) (`Tropics'), and south of \(15^\circ\South\) (`South of Tropics', SH).
The surface KE spectra (dark purple lines) in the extra-tropical hemispheres (NH and SH) broadly follow the same pattern that was found from studying geostrophic velocities \cite{Storer2022}.
Namely, there is a meso-scale peak at \(\approx250~\)km, a NH gyre-peak at \(\approx3\times10^3~\)km, and an ACC peak at \(\approx9\times10^3~\)km.
Our results then extend those previous findings by considering full model velocity, instead of just the geostrophic velocity, thus allowing us to study the entire global ocean including the tropics. 

Consistent with previous studies that focused on regional spectral analyses \cite{Xu2012,khatri2018surface,Uchida2017,Khatri2021,Ajayi2021,Callies2013}, the meso-scale spectral scaling lies between $k^{-5/3}$ and $k^{-3}$ in the global KE spectrum. Due to ageostrophic Ekman flow over the range $10^3~$km to $10^4~$km \cite{Vallis2006}, the extra-tropical gyre-scales in Fig.~\ref{fig:surface_spectra}B-C follow a scaling that is slightly steeper than the $k^{-1}$ previously found for the geostrophic velocity \cite{Storer2022}.
While the extra-tropics exhibit an energy minimum separating the meso-scale and gyre-scale peaks, the tropics, which were not analyzed in \cite{Storer2022}, do not have such a spectral energy minimum. Instead, the tropics yield a global ocean surface KE spectrum that monotonically increases with length scale up to scales of $\approx 10^4~$km.

\subsection*{\MakeUppercase{KE Spectra at Depth}}
Figure~\ref{fig:surface_spectra} provides the first global power spectrum of the ocean as a function of depth (indicated by the colour bar).
At almost all length scales, spectral KE density decreases monotonically with depth \cite{Ajayi2021,Khatri2021}, consistent with the surface intensification of the ocean currents.
However, the magnitude of such decrease in energy with depth is strongly dependent on length scale. A striking feature in Figure~\ref{fig:surface_spectra} is how rapidly gyre-scale energy decays with depth compared to energy at the meso-scales.
For instance, in Figure~\ref{fig:surface_spectra}D, energy density at \(\ell=200~\)km does not decrease noticeably in the upper \(\sim100~\)m, while energy density at \(\ell=10\times10^3~\)km decreases by roughly an order of magnitude over the same depth (see also Fig.~\ref{fig:spectra:SUPP}D in SM). Figure~\ref{fig:surface_spectra}B-C (and Fig.~\ref{fig:spectra:SUPP}C-F in SM) quantifies the gyre-scales' precipitous KE decay with depth. Their KE decreases by a factor of $O(10)$ at $500~$m depth and of $O(100)$ at $2000~$m depth compared to the surface, albeit with notable differences between the NH and SH. 
This surface-trapped gyre-scale motion is consistent, at least in part, with baroclinic Rossby wave adjustment \cite{rintoul2018global} and wind-driven Ekman transport, which is restricted to the Ekman layer in the upper $\sim 100~$m \cite{Vallis2006}.

\subsubsection*{\MakeUppercase{Mesoscales Span the Water Column}}
Outside of the tropics (Figure~\ref{fig:surface_spectra}B-C), the mesoscale spectral peak is present at all depths.
While their spectral energy density decreases by roughly a factor of 15 from the surface to the abyssal ocean, the mesoscales remain energetically dominant, especially considering that energy density at scales larger than \(10^3~\)km decreases by 2--3 orders of magnitude. Figure~\ref{fig:surface_spectra}B-C (and Fig.~\ref{fig:spectra:SUPP}C-F in SM) quantifies the extent to which mesoscales are barotropic \cite{lacasce2020baroclinic}, with their KE decreasing by a factor of $\approx 3$ at $500~$m depth and of $O(10)$ at $2000~$m depth compared to the surface. That the spectral energy \emph{per wavenumber} in the mesoscales is on par with or greater than that of the largest scales underscores the dominance of the mesoscales at all depths.
Oceans are forced at the surface by winds and buoyancy fluxes and, to maintain equilibrium, some of this energy is transferred to the deeper ocean where it is dissipated at the bottom via friction \cite{arbic2009estimates,Nikurashin2011}. 
The mesoscale dominance of the deep ocean spectra highlights their key role in the ocean's energy dissipation pathways.

\subsection*{\MakeUppercase{Global Scale-Transfer of KE}}
\begin{figure*}[tbhp]
    \centering
    \includegraphics[scale=1]{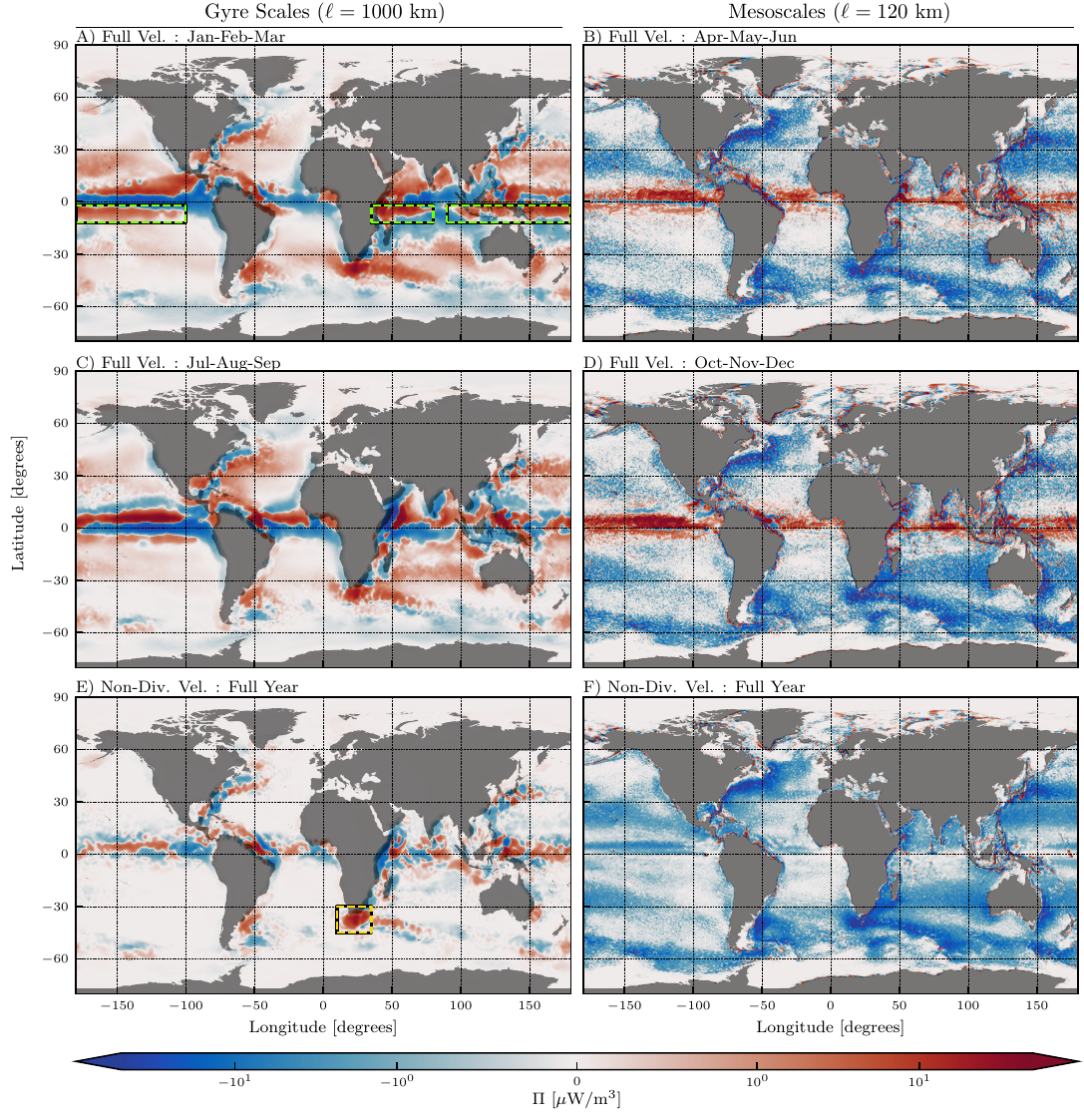}
    \caption{ 
        \textbf{Surface maps of \(\mathbf{\Pi_\ell}\)} at [A,C,E] \(\ell=1000~\)km and [B,D,F] \(\ell=120~\)km from four years (2015--2018), with a positive (negative) value indicating a downscale (upscale) kinetic energy transfer.
        To highlight ITCZ seasonal imprint on gyre-scale KE transfer and seasonal extrema of mesoscale KE transfer, \(\Pi_\ell\), using full velocity, is averaged over
        [A] Jan-Feb-Mar, [B] Apr-May-Jun, [C] Jul-Aug-Sep, and [D] Oct-Nov-Dec. All four seasons are in Figure~\ref{fig:SUPP:maps:full:seasonal} in the SM.
        [E,F] are annual averages of \(\Pi_\ell\) but using only the laterally non-divergent velocity. Seasonal maps equivalent to [E,F] are in Figure~\ref{fig:SUPP:maps:toroidal} in the SM.
        All panels share a common colour scale.
        To preserve scale-dependent symmetries, the coarse-grained flow is allowed to be nonzero within a distance $\ell/2$ beyond land boundaries (grey mask), consistent with the uncertainty principle (see \hyperref[Methods:EkmanVel]{Methods}).
        [A: green dashed boxes] highlight the Southern ITCZ ``red branch'' imprint that occurs during JFM. A corresponding ITCZ imprint can be seen north of the Equator during JAS (\textbf{[C]}) in all basins. 
        [E: orange dashed box] highlights strong downscale KE transfer from geostrophic flow shear between the Agulhas and ACC. 
    }
    \label{fig:SUPP:Pi_maps_1000km}
\end{figure*}

Figure~\ref{fig:SUPP:Pi_maps_1000km} provides the first global maps of the surface KE scale-transfer, \(\Pi_\ell\), across \(\ell=1000~\)km (gyre-scales) and \(\ell=120~\)km (mesoscales), averaged over 2015--2018 (see also Figure~\ref{fig:enter-label} in SM using satellite data).
Positive \(\Pi_\ell\) values (red) indicate a downscale transfer of KE from scales larger than \(\ell\) to those smaller than \(\ell\), while negative \(\Pi_\ell\) values (blue) indicate an upscale transfer across $\ell$. These geographic maps highlight a key advantage of the coarse-graining methodology over traditional approaches using Fourier or structure functions: we can retain spatial information while concurrently diagnosing processes at different scales.
The maps in Figure~\ref{fig:SUPP:Pi_maps_1000km}  allow us to associate the scale-transfer of KE with flow properties in different regions, as we discuss below. 
To highlight seasonal trends, gyre-scale KE transfer maps (panels~\ref{fig:SUPP:Pi_maps_1000km}A,C) are averaged over winter and summer months, while the mesoscale energy transfer across $120~$km are averaged over spring and autumn when scale-transfer across that scale is at an extremum (see Figure~\ref{fig:SUPP:maps:full:seasonal} in SM for all seasons).
Panels~\ref{fig:SUPP:Pi_maps_1000km}E-F show full-year means of the KE scale-transfer but only due to the laterally non-divergent component of the ocean currents, which includes geostrophic motions but excludes the divergent Ekman flow (see Figure~\ref{fig:SUPP:maps:toroidal} in SM for seasonal maps). To preserve physical properties (symmetries) at different scales, the coarse-grained flow is allowed to be nonzero within a distance $\ell/2$ beyond the continental boundary over land \cite{Aluie2018,Storer2022,buzzicotti2023coarse}. Forfeiting exact spatial localization to gain scale information is theoretically inevitable due to the uncertainty principle (see \hyperref[Methods]{Methods}).

\begin{figure*}[tbhp]
    \centering
    \includegraphics{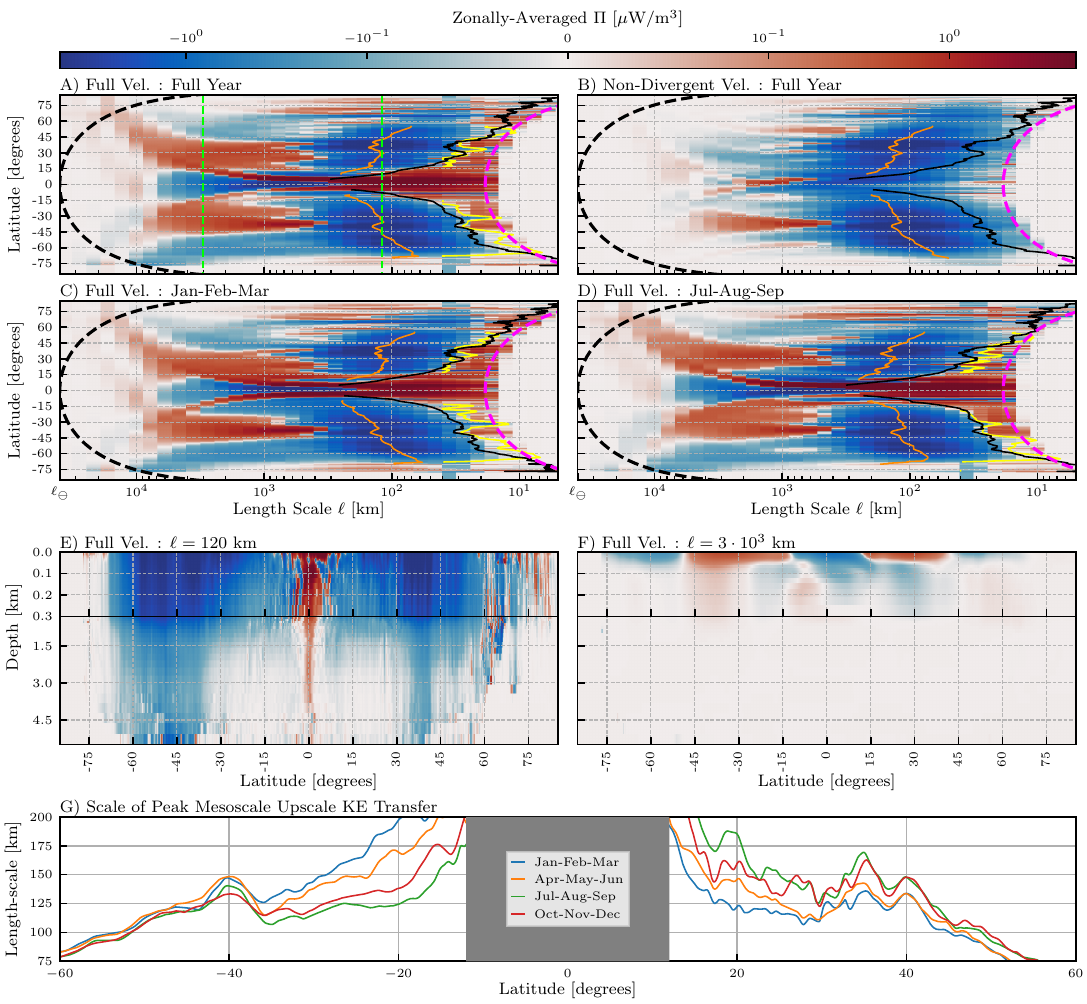}
    \caption{ 
        \textbf{Structure of the KE Scale-Transfer}
        [A-D] The zonally-averaged surface KE scale-transfer as a function of latitude and scale:
        [A] using full velocities and averaged over all months,
        [B] using only laterally non-divergent velocities and averaged over all months,
        [C] using full velocities and averaged over Jan-Feb-Mar, and
        [D] using full velocities and averaged over Jul-Aug-Sep. 
        Negative (positive) values indicate an upscale (downscale) transfer of energy.
        \(\ell_{\ominus}\) denotes the equatorial circumference of the Earth, \(\sim40\times10^3~\)km. 
        [Thick dashed black lines] denote the circumference of a line of constant latitude.
        [Thick dashed pink lines] denote the zonal width of two grid-points.
        [Straight dashed green lines] in [A] denote the length-scales analyzed in [E] and [F].
        In [A-D], [orange lines] show the scale with greatest magnitude upscale KE transfer [``mesoscale peak'', panel G], [yellow lines] show the small-scale transition to downscale transfer, and [black lines] show the length scale at which the Rossby number equals \(0.1\).
        Panels [E,F] show latitude and depth for $\ell=120~$km and $\ell=3\times10^3~$km (dashed green lines in [A]).
        [E,F]: Note that the vertical axis is split at 300~m depth, so that the upper 300~m are emphasized.
        The colour bar is the same for all panels.
        [G] shows the mesoscale peak for each of the seasonal bands, smoothed using a 3-degree moving average.
        Figure~\ref{fig:SUPP:zonal_Pi:annotated} reproduces [C,D] with additional annotations.
    }
    \label{fig:Pi_zonal_means}
\end{figure*}

Figure~\ref{fig:Pi_zonal_means} complements the scale-transfer maps in Figure \ref{fig:SUPP:Pi_maps_1000km} by showing the zonally (east-west) averaged scale-transfer as a function of latitude, length-scale, and depth.
Figure~\ref{fig:Pi_zonal_means}A presents time-mean surface KE scale-transfer as a function of latitude and scale. In the SM, Figure~\ref{fig:SUPP:zonal_Pi:annotated} is an annotated version of Figure~\ref{fig:Pi_zonal_means}, and Figure~\ref{fig:SUPP:AVISO} presents the same analysis from AVISO satellite product.

\subsubsection*{\MakeUppercase{Mesoscale KE Transfer}}
Maps of KE scale-transfer at the mesoscales  (Figure~\ref{fig:SUPP:Pi_maps_1000km}B,D) show the pervasiveness of the upscale KE transfer, which dominates the extratropical global ocean and is consistent with geostrophic turbulence theory \cite{charney1971geostrophic,Vallis2006}. 
The upscale transfer is most intense in dynamically active regions, including the Gulf Stream, Kuroshio, and the ACC, due to strong mesoscale eddy activity and large baroclinic growth rates \cite{tulloch2011scales}.  In contrast, equatorial regions, which are not constrained by geostrophy, show a prominent downscale KE transfer (red) across scale of $120~$km in Figures~\ref{fig:SUPP:Pi_maps_1000km}B,D.
Comparing maps in panels B and D, we find indications that the mesoscale upscale KE transfer across $120~$km is stronger during the local spring.
This behavior is particularly noticeable in the subtropics, such as in the Kuroshio region and around Australia. 
Panel F shows that the laterally non-divergent flow component, which excludes upwelling/downwelling motions, exhibits a more pronounced upscale transfer, including in the tropics. 

In Figure~\ref{fig:Pi_zonal_means}A-D, the upscale transfer (blue) is prominent over a wide range of mesoscales, and spans roughly \([65^\circ\South,15^\circ\South]\) and \([15^\circ\North, 55^\circ\North]\) (see also Figure~\ref{fig:SUPP:AVISO} in SM). In each panel, we outline three length-scales of interest: 1) the length-scale with the strongest mesoscale upscale KE transfer (orange, ``mesoscale peak''), 2) the scale at which the KE transfer transitions to downscale (black, ``downscale transition''), and 3) the length-scale corresponding to Rossby number of \(0.1\) (yellow, \(\ell_{\mathrm{Ro=0.1}}\)), with larger values of Rossby number indicating a reduced influence from Earth's rotation (see \hyperref[Methods:EkmanVel]{Methods}).

The scale of the peak upscale KE transfer (orange lines in Figure~\ref{fig:Pi_zonal_means}A-D, also shown in panel~\ref{fig:Pi_zonal_means}G) 
ranges from $\approx200~$km near the equator to $\approx80~$km near the poles. The peak scale generally decreases polewards except in strong current systems (see panel~\ref{fig:Pi_zonal_means}G and panel~\ref{fig:SUPP:AVISO}C in SM). There is a notable increase in the peak scale at \(40^\circ\South\) associated with the ACC, and two smaller increases at \(\sim35^\circ\North\) and \(40^\circ\North\), associated with the Gulf Stream and Kuroshio. In these strong currents, the mesoscale nonlinear interactions are more intense (Figure~\ref{fig:SUPP:Pi_maps_1000km}B,D), transferring energy upscale over a wider range of scales.
This latitudinal dependence of the length scale corresponding to the peak upscale KE transfer is consistent with previous studies that used Fourier analysis to compute KE scale-transfer \cite{Scott2005,tulloch2011scales,khatri2018surface,Ajayi2021}, with our analysis here covering the global ocean without being confined to regional boxes.

Transition from upscale to downscale KE transfer in Figure~\ref{fig:Pi_zonal_means} tracks \(\ell_{\mathrm{Ro}=0.1}\) relatively well, suggesting that the transition to downscale KE transfer is driven by unbalanced motions, consistent with previous work \cite{Ajayi2021,Balwada2022,Srinivasan2023}. Excluding the irrotational flow, scale-transfer from the laterally non-divergent flow at these small scales seems to exhibit an upscale transfer in Figure~\ref{fig:Pi_zonal_means}B.  Given that the resolution of the dataset used here is \(\sfrac{1}{12}^\circ\) (\(\sim9~\)km at the equator), this transition can only be considered marginally resolved and submesoscale-permitting at best \cite{Uchida2017}, and should be further investigated with higher resolution data.

The tropics are the only latitudes that demonstrate a persistent downscale transfer (red) over mesoscales (\(\ell<500~\)~km). 
As \(\ell\) increases from $\cO(100)~$km to scales $>1000~$km over the tropics in Figure~\ref{fig:Pi_zonal_means}, the downscale energy transfer bifurcates into two intense off-equatorial branches (red) that lie within \(\approx\pm10^\circ\) of the equator (``red branches'' in Figure~\ref{fig:SUPP:zonal_Pi:annotated}). 
The two branches have a distinct north-south asymmetry, with the northern band exhibiting a stronger downscale transfer due to the asymmetric seasonal migration of the Inter-Tropical Convergence Zone (ITCZ) as we elaborate below.

\subsubsection*{\MakeUppercase{Gyre-scale KE Transfer and the\\Atmospheric Imprint}}
Maps of the KE transfer across gyre-scales in Figures~\ref{fig:SUPP:Pi_maps_1000km}A,C and their zonal average in 
Figure~\ref{fig:Pi_zonal_means}A reveal an imprint of the global atmospheric circulation. This imprint is most clear from Figure~\ref{fig:Pi_zonal_means}A, where we find five `cells' of alternating upscale and downscale transfer at the ocean surface (c.f. ``Atmospheric Cell Imprint'' in Figure~\ref{fig:SUPP:zonal_Pi:annotated}). 
Each cell spans \(\approx30^\circ\) in latitude. They are centred at the Equator, horse latitudes ($30^\circ$N and $30^\circ$S), and polar fronts ($60^\circ$N and $60^\circ$S), which mark the transition zones between the atmospheric Hadley, Ferrel, and polar cells.

This atmospheric signature on the gyre-scale oceanic KE scale-transfer can be explained by the wind-driven Ekman transport within those bands \cite{Vallis2006}. 
Zonal surface wind stress \((\tau_\lambda)\) induces meridional (north-south) Ekman velocity (\(u^E_\phi\), see \hyperref[Methods:EkmanVel]{Methods}),
such that a meridionally-alternating wind direction produces  meridionally-alternating divergent and convergent flows within the ocean's Ekman layer (top $\sim100~$m). 
These flows would respectively `stretch' and `compress' oceanic motions, leading to KE transfer to larger and smaller scales, respectively. We shall see below that much of the gyre-scale transfer occurs due to energy exchange with the mesoscales.
A concept worth emphasizing is that energy can undergo an inertial scale-transfer of KE in the presence of coherent convergent or divergent flow structures such as in shocks, rarefaction waves, and fronts \cite{kraichnan1974kolmogorov,Aluie11,Aluie13,buzzicotti2016phase}. While KE scale-transfer due to convergent and divergent (unbalanced) motions has been analyzed at the submesoscales in regional models \cite{schubert2020submesoscale,Srinivasan2023}, the results presented here are the first to show scale-transfer due to the convergent and divergent Ekman flow at gyre-scales.
To reinforce this interpretation, panel~\ref{fig:SUPP:Pi_maps_1000km}E and panel~\ref{fig:Pi_zonal_means}B present \(\Pi_\ell\) arising solely from the laterally non-divergent (toroidal) flow component, which lacks the necessarily divergent Ekman flow component and the associated atmospheric-cell pattern. 
Further support for this interpretation comes from examining the depth profile of the gyre-scale KE transfer (Figure~\ref{fig:Pi_zonal_means}F, discussed in the next paragraph), where we find that the alternating upscale and downscale transfer bands are localized to the top $\approx100~$m in the ocean, which is approximately the same depth as the Ekman layer. 
From panels~\ref{fig:Pi_zonal_means}A,C,D we note a north-south asymmetry in the gyre-scale KE transfer poleward of \(45^\circ\).
South of \(45^\circ\South\), the upscale transfer (blue) persists to larger scales and with higher intensity than its NH counterpart. This behavior is attributed to continental boundaries, which constrain the upscale transfer (blue) north of \(45^\circ\North\) unlike in the Southern Ocean.

\subsubsection*{\MakeUppercase{Depth Profiles of KE Scale-Transfer}}
Figure~\ref{fig:Pi_zonal_means}E-F presents the latitude-depth profiles of \(\Pi_\ell\) for \(\ell=120\)~km and \(\ell=3\times10^3~\)km, which represent the mesoscale and gyre-scale KE scale-transfer signals in Figures~\ref{fig:SUPP:Pi_maps_1000km},\ref{fig:Pi_zonal_means}A.
The mesoscale transfer (panel \ref{fig:Pi_zonal_means}E), in addition to having strong intensity, penetrates the entire water column.
The scale-transfer is surface intensified (note the log-scale in the colour bar), but there is a clear upscale transfer down to \(5~\)km depth, especially between \([60^\circ\South,30^\circ\South]\) and \([30^\circ\North,45^\circ\North]\), which are the approximate latitudes of the ACC, Gulf Stream, and Kuroshio.
This result provides evidence that the upscale mesoscale transfer has a substantial barotropic component, in accord with the theory of geostrophic turbulence \cite{Salmon80}.
In contrast, the gyre-scale KE transfer (panel~\ref{fig:Pi_zonal_means}F) is mostly surface-trapped to the upper \(100~\)m of the ocean, where wind effects are most pronounced. Panel~\ref{fig:Pi_zonal_means}F also demonstrates that the downscale/upscale KE transfer due to Ekman flow convergence/divergence near the surface is not cancelled by the return Ekman flow divergence/convergence at greater depth. This lack of cancellation is because the KE scale-transfer (eq.~\ref{eq:define:Pi} in \hyperref[Methods]{Methods}) arises from flow strain at scales $>\ell$ acting against stress from motions at scales $<\ell$. These subscale motions are much stronger near the surface than at depth (Fig.~\ref{fig:surface_spectra} and SM Fig.~\ref{fig:spectra:SUPP}), underscoring the importance of vertical inhomogeneity in the oceanic scale-transfer of KE.

In the equatorial region, we can see clearly in Panel~\ref{fig:Pi_zonal_means}E that mesoscale KE transfer differs substantially from other latitudes. It is characterized by strong down-scale KE transfer, which penetrates several kilometers into the water column.
This downscale transfer may be related to shear induced by alternating equatorial deep jets \cite{Menesguen2019,Hua2008}, with the meridionally broader but shallower downscale transfer possibly owing to the subsurface counter-currents. Further analysis of this phenomenon is perhaps better investigated using regional modeling of the equatorial region at higher vertical resolutions.

\subsubsection*{\MakeUppercase{Departures from Geostrophic Theory}}
The laterally non-divergent flow, which includes geostroph\-ic (balanced) motions, exhibits a general tendency to transfer KE upscale (blue) at all latitudes and depths for scales $<1000~$km (c.f. Figures~\ref{fig:SUPP:Pi_maps_1000km}F,~\ref{fig:Pi_zonal_means}B). 
This mesoscale upscale KE transfer is consistent with idealized geostrophic turbulence theory \cite{Salmon80,Vallis2006}, which neglects non-ideal effects from winds, regional inhomogeneity and boundaries.
As we find here, these non-ideal effects, which exist in the realistic NEMO global ocean reanalysis and AVISO satellite data, can become important at \emph{gyre-scales}. Geostrophic flow at the gyre-scales (Figure~\ref{fig:SUPP:Pi_maps_1000km}E and Figure~\ref{fig:enter-label}A) exhibits downscale transfer (red) in regions of strong shear, including the flanks of the Gulf Stream and Kuroshio (see also \cite{Aluie2018}). This behavior is perhaps clearest in the Agulhas retroflection (orange box in panel~\ref{fig:SUPP:Pi_maps_1000km}E), where the westward Agulhas current turns back on itself due to strong shear from the eastward flowing ACC. The effect can also be seen from the zonally averaged scale-transfer in Figure~\ref{fig:Pi_zonal_means}B, for \(\ell>500~\)km between (\(45^\circ\South,30^\circ\South\)), where the laterally non-divergent velocity field produces a local net downscale KE transfer.

\begin{figure}
    \centering
    \includegraphics[scale=1]{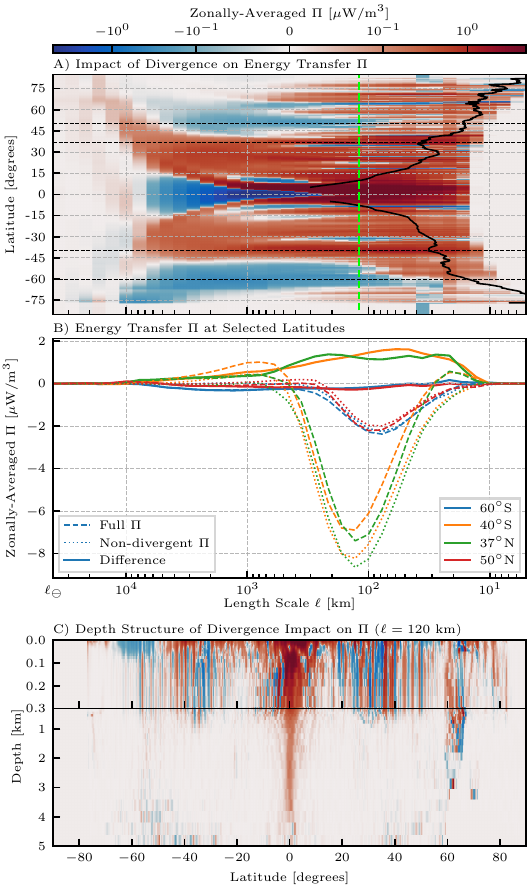}
    \caption{ 
        \textbf{Impact of Divergent Flow on Energy Transfer}
        This extends Figure~\ref{fig:Pi_zonal_means} by highlighting the impact of laterally divergent motions on KE transfer at all scales.
        Panel [A] subtracts \ref{fig:Pi_zonal_means}B from \ref{fig:Pi_zonal_means}A to exclude contributions from self-interactions among non-divergent (geostrophic) flow, thereby showing KE transfer involving interactions with the laterally divergent flow.  Panel [B] shows transfer at select latitudes indicated by horizontal dashed black lines in [A].
        Panel [C] is analogous to \ref{fig:Pi_zonal_means}E.
        The [black contour line in A] is the same as in \ref{fig:Pi_zonal_means}A and shows the length-scale at which the Rossby number is 0.1.
        The [green dashed line in A] shows the length scale used in panel [C].
    }
    \label{fig:DivergenceImpactOnPi}
\end{figure}

Direct evidence for the existence of energy exchange between gyre-scales and mesoscales can be gleaned from Figure~\ref{fig:DivergenceImpactOnPi}. It shows the KE scale-transfer involving interactions with the laterally divergent flow, which includes the gyre-scale Ekman transport. Such KE scale-transfer excludes interactions of the geostrophic (laterally non-divergent) flow with itself shown in Figure~\ref{fig:Pi_zonal_means}B, which is generally characterized by a dominant upscale KE transfer. We emphasize that this Ekman-induced scale-transfer is measured directly from the ocean currents, and not indirectly inferred from the wind stress curl.
Figure~\ref{fig:DivergenceImpactOnPi}A can be obtained simply as the difference of Figure~\ref{fig:Pi_zonal_means}A and Figure~\ref{fig:Pi_zonal_means}B, which is possible because the energy transfer term \(\Pi\) (eq.~\ref{eq:define:Pi} in \hyperref[Methods]{Methods}) can be decomposed exactly into contributions from the laterally divergent/non-divergent flow and their nonlinear interactions (see \hyperref[Methods]{Methods}).

After removing the geostrophic upscale cascade (Figure~\ref{fig:Pi_zonal_means}B), Figure~\ref{fig:DivergenceImpactOnPi} reveals that the alternating bands of Ekman-induced energy transfer extend from the gyre-scale to the mesoscales. 
The persistent scale-transfer of energy spanning the range of scales \(\mathcal{O}(10^4)~\)km to \(\mathcal{O}(10^2)~\)km and smaller in Figure~\ref{fig:DivergenceImpactOnPi}A-B demonstrates the ability of gyre-scale motions to exchange KE with the mesoscales. 
At gyre-scales ($>1000~$km), the flow experiences an effective stress from sub-gyre-scale motions (eq.~\ref{eq:define:Pi} in \hyperref[Methods]{Methods}), which is mainly due to the mesoscales ($<500~$km) where most of the energy resides. 
A convergent gyre-scale flow in the subtropics amplifies the KE of mesoscale eddies, much like a piston pushing (adiabatically) on a gas amplifies the KE of gas molecules (i.e. gas temperature or internal energy). 
Energy transfer occurs because a converging piston does work against the stress (pressure) exerted by molecules via pressure-dilatation \((-P\,\mathrm{div}(\mathbf{u}))\), which appears in the internal energy budget of a compressible flow \cite{Aluie13}. Similarly, a divergent gyre-scale flow at latitudes \([45^\circ,70^\circ]\)  attenuates the KE of mesoscale eddies, much like a piston rarefying a gas attenuates the KE of gas molecules.
Such an analogy between mesoscale eddies and gas molecules can be justified thanks to the seeming separation of scales between the gyre-scale spectral peak and the mesoscale peak shown in \cite{Storer2022} and also in Figures~\ref{fig:surface_spectra}B-C. In analogy with the gas molecules' KE increase, gyre-scale compression brings mesoscale eddies closer together and leads to a gain in mesoscale KE due to enhanced self-advection \cite{Onsager49}. Given the wide scale separation, gyre-scale convergence is probably inefficient at amplifying the mesoscales by vortex stretching. 
We present this novel piston-pressure framework not as an established fact, but as a proposed theory for how the gyre-scales and mesoscales can interact in a manner that explains the energy transfer measured in Figure~\ref{fig:DivergenceImpactOnPi}.

Figure~\ref{fig:DivergenceImpactOnPi}B shows that the gyre-scale -- mesoscale transfer due to gyre-scale convergence/divergence at select latitudes is a significant fraction of the upscale mesoscale cascade. When these different processes are combined into the full scale-transfer in Figure~\ref{fig:Pi_zonal_means}A, the upscale mesoscale cascade masks the gyre-scale -- mesoscale transfer.

Figure~\ref{fig:DivergenceImpactOnPi}C examines the depth profile of KE transfer shown in Figure~\ref{fig:DivergenceImpactOnPi}A at $120~$km. 
In the extra-tropics, we see that the alternating latitudinal bands of KE transfer, while being primarily localized to the upper $\approx 100~$m, exhibit columnar features that penetrate to $\approx 300~$m depth. This is similar to the depth over which the upscale mesoscale cascade is strongest in Figure~\ref{fig:Pi_zonal_means}E. 


\subsection*{\MakeUppercase{Seasonality}}
We now report on seasonal variations in the KE scale-transfer at both gyre-scales and mesoscales.

\subsubsection*{\MakeUppercase{Gyrescales - Atmospheric Cells}}
There are three prominent seasonal trends in the gyre-scale (\(1-10\times10^3~\)km) transfer of KE in Figures~\ref{fig:SUPP:Pi_maps_1000km}A,C~and~\ref{fig:Pi_zonal_means}C,D.
First, consider the five latitudinal bands associated with the atmospheric cells (c.f. ``Atmospheric Cell Imprint'' in Figure~\ref{fig:SUPP:zonal_Pi:annotated}), which are known to strengthen and shift equatorward during the winter of each hemisphere \cite{CliftPlumb2008monsoon}. 
This seasonality is clearest in the zonally-averaged scale-transfer in Figure~\ref{fig:Pi_zonal_means}C-D, which shows that the two extra-tropical bands at scales $>1000~$km (one red and one blue) shift equatorward by a few degrees during the winter of each hemisphere. 
Note also (panels~\ref{fig:SUPP:Pi_maps_1000km}A,C) the seasonality of KE transfer associated with the Indian monsoon winds, which have a southward/northward component during winter/summer \cite{CliftPlumb2008monsoon}. 
The resultant Ekman ocean flow (\hyperref[eq:EkmanZonalVel]{eq.~\ref{eq:EkmanZonalVel}} in \hyperref[Methods:EkmanVel]{Methods}) is westward/eastward, which yields a downscale KE transfer (red) off the eastern coasts of the Arabian peninsula, Horn of Africa, and Indian subcontinent in winter due to convergence toward land and an upscale KE transfer (blue) during summer due to divergence away from land.

\subsubsection*{\MakeUppercase{Gyrescales - ``Red Branches'' and the ITCZ Imprint}}
The second prominent seasonal trend is the feature that we term ``red branches'' (c.f. Figure~\ref{fig:SUPP:zonal_Pi:annotated}).
Figure~\ref{fig:Pi_zonal_means}C-D presents two large-scale off-equatorial downscale signals: one equatorward of \(10^\circ\North\), which is present for much of the year but is strongest during Jul-Aug-Sep (JAS), another  equatorward of \(10^\circ\South\) that is present during Jan-Feb-Mar (JFM). 
The red branches are caused by an interplay between the ITCZ and Ekman transport as we shall now explain.

The ITCZ is a latitudinal band at which the northeast and southeast trade winds from the two Hadley cells converge near the equator and yield intense tropical rainfall within \(10^\circ\South\)--\(10^\circ\North\) \cite{schneider2014migrations,kang2018extratropical}. 
In the zonal average, the ITCZ is located in the NH for most of the year but is strongest during the boreal summer, typically shifting to the SH only during the austral summer \cite{mitchell1992annual,philander1996itcz}.
The red branches in Figure~\ref{fig:Pi_zonal_means}C-D at scales $\cO(10^3)~$km align with the ITCZ latitudes \cite{schneider2014migrations}. These red branch latitudes can also be seen from the scale-transfer maps in panels~\ref{fig:SUPP:Pi_maps_1000km}A,C, which show intense downscale KE transfer (red) just off of the equator. During the SH summer, we see a red latitude band (green box in panel~\ref{fig:SUPP:Pi_maps_1000km}A) just south of the equatorial Pacific and Indian oceans. During the NH summer (panel~\ref{fig:SUPP:Pi_maps_1000km}C), an intense downscale transfer band can be seen just north of the equator in the Pacific, Atlantic, and Indian Oceans.

\begin{figure}[tbhp]
   \centering
   \includegraphics{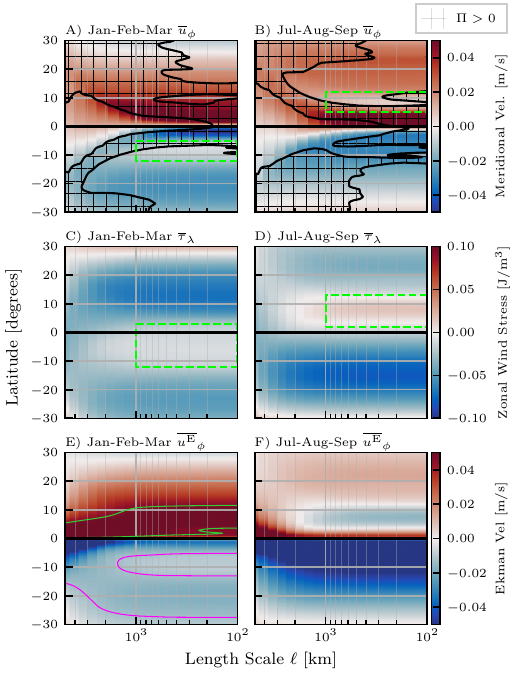}
   \caption{ 
        \textbf{ITCZ Scale-Transfer Mechanism}
        Zonal-means of [A,B] meridional surface velocity (\(u_\phi\)) from NEMO, [C,D] zonal wind stress (\(\tau_\lambda\)), and [E,F] associated meridional Ekman velocity \(u_\phi^E\) (\hyperref[eq:EkmanMeridionalVel]{eq.~\ref{eq:EkmanMeridionalVel}}) within \(30^\circ\) of the equator for filter scales larger than 100~km.
        In [A,B], cross-hatching indicates latitudes at which the zonal-mean surface KE scale-transfer \(\Pi_\ell\) is positive (i.e. downscale).
        Green boxes are included to highlight regions of weak meridional transport and zonal wind stress.
        [E:green lines] show the \(4.7~\mathrm{cm}/\mathrm{s}\) contour, and [E:magenta lines] show the \(-1.2~\mathrm{cm}/\mathrm{s}\) contour.
        Wind data is from ERA5 reanalysis wind components used in the NEMO ocean reanalysis.
   }
   \label{fig:SUPP:ITCZ}
\end{figure}

The ITCZ, being a band of weak zonal winds, is associated with a sudden slowdown in the poleward Ekman flow leaving the Equator (green boxes in Figure~\ref{fig:SUPP:ITCZ}A-B). Such a slowdown is akin to a hydraulic jump in a river encountering an obstacle, causing surface flow convergence and downwelling.
Figure~\ref{fig:SUPP:ITCZ}A-B shows the zonally averaged meridional velocity from NEMO as a function of latitude and scale during JFM (panel~\ref{fig:SUPP:ITCZ}A) and JAS (panel~\ref{fig:SUPP:ITCZ}B). Panels~\ref{fig:SUPP:ITCZ}A-B overlay the same downscale branches (cross-hatching) seen in Figure~\ref{fig:Pi_zonal_means}C-D (red branches).
In the SH, the poleward flow weakens between \(5^\circ\South\)-\(10^\circ\South\) during JFM (green box in panel~\ref{fig:SUPP:ITCZ}A), with a similar weakening present in the NH during JAS (green box in panel~\ref{fig:SUPP:ITCZ}B).
Slowdown in the poleward flow is caused by the zonal relative wind stress in panels~\ref{fig:SUPP:ITCZ}C-D, which show that the typically-westward wind stress weakens in the ITCZ band.
Recall that meridional Ekman velocity, $u^E_\phi \propto - \tau_\lambda/f$ (\hyperref[eq:EkmanMeridionalVel]{eq.~\ref{eq:EkmanMeridionalVel}} in \hyperref[Methods:EkmanVel]{Methods}), is proportional to the zonal wind stress, $\tau_\lambda$, 
so that the change in sign of the Coriolis parameter, \(f\), across the equator induces divergence under a westward wind, and the small magnitude of \(f\) near the equator means that even modest wind stress can produce a substantial flow.
The Ekman flow associated with the wind stress is shown in panels~\ref{fig:SUPP:ITCZ}E-F.

We can also explain why the NH red branch is still present during JFM (Figure~\ref{fig:Pi_zonal_means}C), when the ITCZ is mostly south of the Equator. The northern branch is perennial because the ITCZ (zonally averaged) does not migrate as far south in the boreal winter as it migrates north in the boreal summer, maintaining an oceanic imprint north of the Equator.
Panel~\ref{fig:SUPP:ITCZ}E shows contours of weak Ekman flow: \(10~\mathrm{cm}/\mathrm{s}\) (green) and \(-2.5~\mathrm{cm}/\mathrm{s}\) (magenta). 
During JFM, the NH poleward Ekman flow leaving the Equator (panel~\ref{fig:SUPP:ITCZ}E) undergoes an acceleration followed by sudden slowdown (green contour lines at the equator and at $\approx5^\circ\North$), resulting in a convergent flow and a downscale KE transfer.
Note that exactly at the Equator, we always have an upscale KE transfer at gyre-scales (blue in Figure~\ref{fig:Pi_zonal_means}). This transfer is due to Ekman flow divergence from a sharp southward to northward flow reversal at the Equator caused by a reversal in the direction of the Coriolis force (\hyperref[eq:EkmanMeridionalVel]{eq.~7} in \hyperref[Methods:EkmanVel]{Methods}).

\subsubsection*{\MakeUppercase{Gyrescales - ``Blue Tongue''}}
The third gyre-scale seasonal trend is also caused by the ITCZ, albeit indirectly, appearing during a hemisphere's summer (Figure~\ref{fig:Pi_zonal_means}C-D) as an upscale `tongue' (blue). It emerges at scales larger than $500~$km in the subtropics, poleward of the ITCZ's ``red branch'' ocean imprint, between latitudes \(5^\circ\) and \(20^\circ\) (c.f. ``blue tongue'' in Figure~\ref{fig:SUPP:zonal_Pi:annotated}). From the map in  panel~\ref{fig:SUPP:Pi_maps_1000km}A, we see that during JFM, the blue tongue is mostly due to an upscale transfer in the Indian Ocean, just south of the downscale (red) band, extending from Madagascar to Australia. During JAS, panel~\ref{fig:SUPP:Pi_maps_1000km}C shows that the blue tongue in the NH is due to an upscale KE transfer that is prominent in the northern tropics of the Atlantic and Pacific. 

Blue tongue regions are caused by the Ekman flow's poleward reacceleration after encountering the slowdown caused by the ITCZ. In Figure~\ref{fig:SUPP:ITCZ}, we can see that poleward of the slowdown (green boxes in panels ~\ref{fig:SUPP:ITCZ}A-B), the flow speed increases. This behavior can also be understood from magneta contours of the Ekman flow in panel~\ref{fig:SUPP:ITCZ}E, where the southward flow at $\approx10^\circ\South$ increases in speed before slowing down again at $\approx30^\circ\South$. Such reacceleration between $10^\circ\South$-$20^\circ\South$ is associated with a divergence and Ekman upwelling, manifested as a summer blue tongue in panels~\ref{fig:Pi_zonal_means}C-D.

\subsubsection*{\MakeUppercase{Mesoscale Cascade Peak}}
The mesoscale cascade is qualitatively consistent across the four seasons. 
Figure~\ref{fig:Pi_zonal_means}G shows the length-scale with the strongest mesoscale inverse cascade as a function of latitude for each season. 
While the seasonal trends are clearer in the southern hemisphere, we can see that in both hemispheres the mesoscale cascade peak scale is largest during the local summer and exhibits the greatest seasonal variation in the subtropics, between \(15^\circ\) and \(30^\circ\).

\subsubsection*{\MakeUppercase{Seasonality of Mesoscale KE Spectrum and the Cascade}}
\begin{figure}[tbhp]
    \centering
    \includegraphics[width=8.66cm]{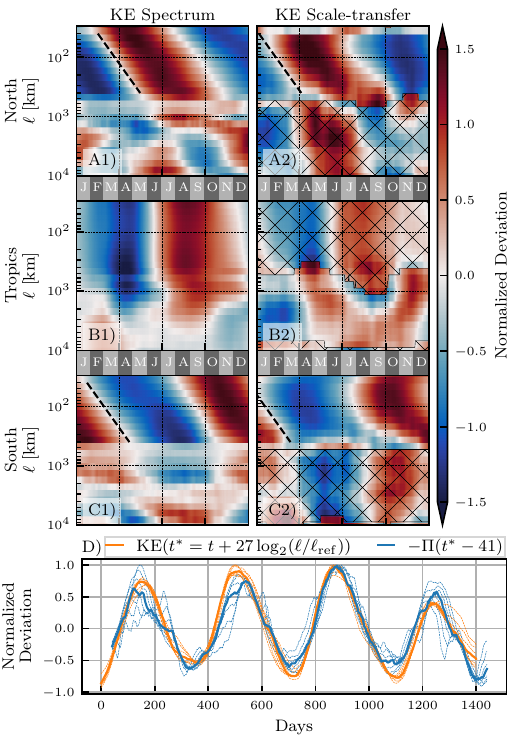}
    \caption{ 
        \textbf{Spectral Advection of Surface KE and \(\mathbf{\Pi_\ell}\).}
        For each scale, panels show normalized variation (z-score) of the time-series of [A1,B1,C1] the KE filtering spectrum, $\OL{E}(k_\ell)$ as in \cite{Storer2022} and [A2,B2,C2] the KE scale-transfer \(\Pi_\ell\) .
        A 60-day moving average is applied to remove high-frequency signals \cite{Storer2022}.
        For KE, red indicates where KE is higher than the time mean.
        For \(\Pi_\ell\), red indicates where the \emph{magnitude} of the scale-transfer is higher than the time mean, with hatching indicating where the scale-transfer is downscale.
        The full time-series has been averaged onto a single `typical' year profile.
        [Dashed black lines] indicate a spectral advection speed of 27 days per octave.
        In [A2,B2,C2] showing \(\Pi_\ell\), the  dashed black lines are plotted 41 days earlier in the year than the corresponding lines in [A1,B1,C1] showing $\OL{E}(k_\ell)$.
        [Panel D] shows $\OL{E}(k_\ell)$ (orange) and \(-\Pi_\ell\) (blue) for length-scales between 60~km and 400~km in the NH.
        Both $\OL{E}(k_\ell)$ and \(-\Pi_\ell\) have been shifted by 27 days per octave relative to 77~km.
        \(-\Pi_\ell\) was then phase-shifted an additional 41~days.
        The reference scale (\(\ell_{\mathrm{ref}}=77~\)km) is an arbitrary reference to show scale self-similarity, as all plots collapse onto the same curve.
        [Thin dashed lines] in [D] correspond to individual scales, while [thick solid lines] show the median across those scales.
    }
    \label{fig:seasonality_contours}
\end{figure}

Figure~\ref{fig:seasonality_contours}A-C shows the seasonality of the surface KE spectrum, $\OL{E}(k_\ell)$, and scale-transfer, \(\Pi_\ell\), as a function of time and scale.
Similar to \cite{Storer2022}, we show that outside of the Tropics and within the scale-range of 40--400~km, larger scales reach their seasonal KE maximum later than smaller scales (panels~\ref{fig:seasonality_contours}A1-C1).
This delay can be regarded as a `spectral advection' signal in which energy moves to larger scales with time, at a time-scale of \(\approx27~\)days for a two-fold increase in \(\ell\).
The spectral advection speed is illustrated by the dashed black lines in Figure~\ref{fig:seasonality_contours}A1,C1. 
This speed is slightly faster than what was found in \cite{Storer2022}, which was 35--45 days per octave for sea surface height derived geostrophic velocities and may be considered as an indication that the geostrophic flow has a higher inertia in responding to the seasonal changes in atmospheric forcing.
Within the Tropics, there is no discernible spectral advection signal, and instead all sub-gyre scales (smaller than 1000~km) reach their seasonal KE maximum at the same time. 

From Figure~\ref{fig:seasonality_contours}{A2-C2}, we see that the cascade has spectral advection signal similar to that of KE within the scale-band of 60--400~km in both hemispheres. Applying the same analysis shows again a period of  $\approx27~$days per octave. While our results are consistent with previous work \cite{sasaki2014impact,callies2015seasonality,Khatri2021,Balwada2022} reporting on the mesoscale seasonality, prior focus had often been on seasonality of the spectral power-law scaling in small regions using Fourier analysis. Those studies did not report on the spectral advection shown in Figure~\ref{fig:seasonality_contours} for the global ocean.

\subsubsection*{\MakeUppercase{Time Lag between KE Spectrum and KE Cascade}}
For any given scale, the energy cascade \(\Pi_\ell\) reaches its seasonal maximum before the KE spectrum $\OL{E}(k_\ell)$ does, which lags by approximately 41 days.
This time lag is highlighted in Figure~\ref{fig:seasonality_contours} in two ways.
First, the reference dashed black lines in panels~\ref{fig:seasonality_contours}A2,C2, which show the cascade's spectral advection speed in the NH and SH, are shifted 41 days earlier than the corresponding lines in panels~\ref{fig:seasonality_contours}A1,C1, which show the spectrum's spectral advection speed. This result shows that at any scale $\ell$, \(\Pi_\ell\) and $\OL{E}(k_\ell)$ are 41 days out of phase in their seasonal cycle.
Second, in Figure~\ref{fig:seasonality_contours}D we plot the normalized variation of the spectrum and cascade for several scales between 60~km and 400~km.
Plots of $\OL{E}(k_\ell)$ (orange) and \(\Pi_\ell\) (blue) are time-shifted, with the spectrum plotted as a function of \(t^* = t + 27\log_2(\sfrac{\ell}{\ell_{\mathrm{ref}}})\) and the cascade as a function of \(t^\dagger = t^* - 41\).
Note that \(\ell_{\mathrm{ref}}=77~\)km is an arbitrary ``reference scale.'' Changing \(\ell_{\mathrm{ref}}\) merely yields a uniform time-shift of all plots in Figure~\ref{fig:seasonality_contours}D. 
With these time-shifts, the seasonal signals for $\OL{E}(k_\ell)$ and \(\Pi_\ell\) (blue and orange  in panel~\ref{fig:seasonality_contours}D) collapse on each other for all plotted scales between 60~km and 400~km. This result is evidence of self-similar dynamics in the mesoscale range, whereby temporal evolution at different length-scales appears identical when properly rescaled. That the cascade peak precedes a peak in the KE spectrum at different scales in Figure~\ref{fig:seasonality_contours} is suggestive of a causal relation between the cascade and variations in the energy content at different mesoscales.

\subsubsection*{\MakeUppercase{Scale-locality of the Mesoscale Cascade}}
 The cascade $\Pi_\ell$ we are measuring quantifies the rate of KE being transferred across scale $\ell$ and whether it is upscale or downscale. The above findings allow us to make an important conclusion about the length-scales at which the cascade is depositing energy. Using two key results from Figure~\ref{fig:seasonality_contours}: (1) the spectral advection for both \(\Pi_\ell\) and  $\OL{E}(k)$ is 27 days per octave, and (2) the time-lag between \(\Pi_\ell\) and $\OL{E}(k)$ is 41 days, we can infer that \(\Pi_\ell\) at scale $\ell$ is in-phase with $\frac{d}{dt}\OL{E}(k)$, the tendency of the KE spectrum at scale $k^{-1}=3.6\,\ell$ (see \hyperref[Methods]{Methods}). This correlation suggests that energy being transferred across scale $\ell$ is primarily deposited at scale $\approx4\,\ell$, which is in agreement with predictions from 2D turbulence theory \cite{chen2006physical}. We caution, however, that unlike in idealized turbulence where the system is closed, in the ocean there are other possible energy sources/sinks besides the cascade that may influence the KE seasonal cycle of scales. While agreement with ideal 2D turbulence theory suggests that  the cascade is probably the dominant energy source for this range of scales, ascertaining it requires probing other energy pathways, including potential energy release \cite{Vallis2006,loose2023diagnosing} and eddy-damping \cite{Zhai2008,Renault2018,Rai2021}.

\subsection*{\MakeUppercase{Cascade Through the Global Ocean Volume}}
\begin{figure}[tbhp]
    \centering
    \includegraphics[width=8.66cm]{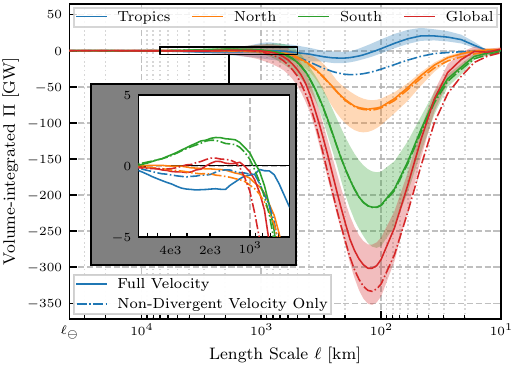}
    \caption{ 
        \textbf{Kinetic Energy Cascade}
        Volume-integrated \(\Pi_\ell\) [GigaWatts] over the global ocean, north of \(15^\circ\North\), between \(15^\circ\South\) and \(15^\circ\North\), and south of \(15^\circ\South\).
        \(\Pi_\ell<0\) signals an upscale transfer whereas \(\Pi_\ell>0\) is downscale.
        Lines correspond to the time-median value with envelopes showing the full temporal range.
        \(\ell_{\ominus}\) denotes the equatorial circumference of the Earth, \(\sim40\times10^3~\)km. 
        Note that the horizontal axis, \(\ell\), decreases to the right (so that the filtering wavenumber, $\ell^{-1}$, increases to the right). 
        The in-set axis zooms in on the plot portion outlined in the black box, and has the same units as the main axes.
        Solid lines correspond to results using the full velocity, while dash-dotted lines correspond to results obtained using only the toroidal (laterally non-divergent) flow component.
        In the in-set axes, 2e3,4e3 are short-hand for \(2\times10^3\) and \(4\times10^3\) respectively.
    }
    \label{fig:Volume_Integrated_Pi}
\end{figure}

Figure~\ref{fig:Volume_Integrated_Pi} presents the scale-transfer of KE \((\Pi_\ell)\) volume-integrated over the global ocean and three regions of interest. 
Outside of the Tropics, the dominant behaviour is a strong upscale cascade (\(\Pi_\ell<0\)) over the mesoscales.
The upscale energy transfer peaks at \(\ell\approx125~\)km, with a total energy transfer rate of \(\approx300~\)GigaWatts [GW]. 
This is the first estimate for the global ocean KE cascade. 
We emphasize that no such estimate is available using either Fourier analysis in regional boxes or  structure functions.
For comparison, this transfer rate of \(\approx300~\)GW is nearly three times larger than a previous global estimate by \cite{VonStorch2012} using a temporal Reynolds decomposition, which showed that \(110~\)GW is transferred from the time-mean to the time-varying flow.
It is also roughly 40\% of the estimated 760~GW that is transferred into the geostrophic surface currents from wind \cite{Hughes2008,scott2009update} and, like the wind input, is dominated by the southern hemisphere. 
Excluding the laterally divergent (ageostrophic) flow, which is mostly in the tropics, the upscale mesoscale cascade is slightly stronger at \(\approx325~\)GW.
Recently, \cite{Rai2021} showed that while wind forcing drives scales larger than 260~km, it on average removes energy from scales smaller than 260~km through eddy damping. Our result here implies that a majority of the mesoscale cascade, including the peak, occurs on length-scales that are on average damped by the winds.

Fig.~\ref{fig:Volume_Integrated_Pi} is simply a meridionally integrated version of Fig.~\ref{fig:Pi_zonal_means}. From Fig.~\ref{fig:Pi_zonal_means}, it can be seen that $120~$km is not the peak transfer scale across all latitudes, owing to variations in the Coriolis parameter. It may be tempting to spatially integrate the scale-transfer $\Pi_{\ell(\phi)}$ across different scales $\ell(\phi)$, which vary with latitude $\phi$ rather than retaining a fixed $\ell$ before spatially integrating as we do in Fig.~\ref{fig:Volume_Integrated_Pi}. However, using a filtering kernel of varying width $\ell(\phi)$ to coarse-grain the governing equations does not yield a term $\Pi_{\ell(\phi)}$ in the energy budget because such filtering does not commute with derivatives (see \hyperref[Methods]{Methods}), which would then render a spatially integrated  $\Pi_{\ell(\phi)}$ of little dynamical meaning.  An advantage of our scale-decomposition is that it guarantees energy conservation \cite{buzzicotti2023coarse}, which follows from our generalized convolution commuting with spatial derivatives on the sphere, thereby preserving scale-dependent symmetries \cite{Aluie2019}.

\subsubsection*{\MakeUppercase{Southern Hemisphere Dominance of the \\ Mesoscale Cascade}}
The majority of the upscale transfer occurs south of the tropics (Figure~\ref{fig:Volume_Integrated_Pi}), with a peak SH cascade value approximately 2.7 times larger than the NH value. 
Since the total water volume of the SH (\(\approx45\%\) of the global ocean volume) is roughly 1.7 times larger than the NH (\(\approx26\%\) of the ocean), the discrepancy in energy cascade cannot be solely attributed to increased volume, and instead indicates higher mean \(\Pi_\ell\) density.
Since the net energy cascade in the Tropics is an order of magnitude smaller, the global mesoscale energy cascade roughly partitions as \(73\%\) occurring in SH and \(27\%\) in NH.

\subsubsection*{\MakeUppercase{Scale-Transfer in the Tropics}}
The tropics (\(15^\circ\South\)--\(15^\circ\North\)) present a qualitatively different \(\Pi\) signature than the extra-tropics (Figure~\ref{fig:Volume_Integrated_Pi}).
We find that the volume-integrated \(\Pi_\ell\) is an order of magnitude smaller than that of the extra-tropics, despite having a volume slightly larger than NH at \(\approx29\%\) of the total ocean volume. 
More importantly, the tropical KE transfer is \emph{downscale} for length-scales smaller than \(\sim100~\)km. Excluding the divergent (ageostrophic) flow, the transfer in the tropics is upscale.
This result is consistent with the notion that flow in the tropics is less constrained by 2D-like geostrophic dynamics and is more similar to 3-dimensional turbulence, which exhibits a downscale cascade \cite{Vallis2006,AlexakisBiferale}.

\subsubsection*{\MakeUppercase{Gyre-scale KE Scale-Transfer}}
The in-set axes of Figure~\ref{fig:Volume_Integrated_Pi} focuses on the range of \(500\)~km to \(7\times10^3\)~km. Being surface-trapped and exhibiting alternating upscale and downscale transfer with latitude, the gyre-scale signal has a much smaller volume-integrated magnitude ($\cO(1)~$GW) relative to the mesoscale cascade, which is mostly upscale and penetrates the entire water column (Fig.~\ref{fig:Pi_zonal_means}E-F). Despite being much weaker on a global spatial average, the gyre-scale transfer is spatio-temporally persistent and directly affects the globally coherent oceanic circulation patterns.
South of the tropics, these scales exhibit a net downscale transfer even when removing the Ekman flow component (non-divergent green plot in Figure~\ref{fig:Volume_Integrated_Pi}, inset). We have seen from the maps in panel~\ref{fig:SUPP:Pi_maps_1000km}E that such downscale transfer appears in regions of strong shear, most prominently between the eastward ACC and the westward Agulhas current (see also Fig.~\ref{fig:Pi_zonal_means}B).  
This result presents another north/south asymmetry, as the NH exhibits a net upscale transfer over the gyre-scales (orange plot in Figure~\ref{fig:Volume_Integrated_Pi}, inset). While there is a downscale transfer in the flanks of the Gulf Stream and Kuroshio currents (red in panel~\ref{fig:SUPP:Pi_maps_1000km}E), the shear is not sufficiently strong to dominate over the upscale transfer in the currents' cores. Flow in the tropics exhibits an upscale transfer on average (blue in Figure~\ref{fig:Volume_Integrated_Pi}, inset), primarily due to the equatorial Ekman divergence at gyre-scales (see maps in Fig.~\ref{fig:SUPP:Pi_maps_1000km}). Yet, even if the divergent component is excluded, an upscale KE transfer is still present at those scales  (dashed blue in Figure~\ref{fig:Volume_Integrated_Pi}, inset).

\section*{\MakeUppercase{Discussion}}
Using a coarse-graining approach developed for analyzing scale dynamics on the sphere \cite{Aluie2019}, and implementing it in an efficient parallel code, \textit{FlowSieve} \cite{Storer_FlowSieve}, we were able to chart the first global maps of the KE cascade in the ocean. Scale-transfer of oceanic KE across gyre-scales revealed a prominent imprint of the atmosphere's Hadley, Ferrel, and polar circulation cells through five latitudinal bands of alternating upscale and downscale KE transfer. We provided evidence that such gyre-scale transfer occurs due to KE exchange with the mesoscales, and proposed a theory to explain the transfer mechanism. Our analysis also found that the atmosphere's ITCZ produces a narrow latitudinal band of intense downscale transfer, which exhibits a seasonal meridional shift following the ITCZ. 

An important accomplishment of this work is quantifying the Ekman transport's role in the KE scale-transfer directly from oceanic velocity. Traditionally, Ekman transport had been inferred only indirectly from wind stress. This was key to demonstrating the existence of energy exchange between gyre-scales and mesoscales, which can be understood using a piston-pressure framework that is unorthodox in physical oceanography. This transfer is not part of the standard theory of gyre circulation, either Stommel's or Fofonoff's \cite{Vallis2006}. The existence of such transfer implies that the mesoscale eddies play an important role in the momentum of gyre-scale by exerting an effective pressure (normal stress). This was not recognized before, probably because typical analysis of the gyre circulation relies on Sverdrup-like relations based on vorticity balance \cite{Vallis2006}, which, being the curl of momentum balance, misses the effective mesoscale pressure process underlying the gyre-scales -- mesoscales transfer.

Probing the scale-transfer of KE across scales ranging from 10~km to $40\times10^3~$km, we were also able to provide the first estimate of the global oceanic cascade, which has a peak upscale transfer rate of $300~$GW. 
This is a substantial fraction ($\approx\sfrac{1}{3}$) of the wind power driving the oceanic general circulation \cite{Hughes2008,scott2009update} and constitutes a previously unquantified source in the global energy budget of the ocean's mesoscales between $100~$km and $500~$km in size \cite{Ferrari2009}.
The comprehensive scale analysis undertaken in this work, spanning 3.5 decades in length-scale, 4 years of daily-averaged velocity, and covering the global ocean volume, including 50 depth levels, has not been performed before. 
In Fig.~\ref{fig:SUPP:AVISO} in the SM, we also present supporting analysis from 9-years of observational data from satellite altimetry. 

We showed that both the mesoscale upscale KE cascade and KE spectrum display a type of scale self-similarity going to larger scales until reaching $\approx500~$km to $10^3~$km. Our analysis indicates that the mesoscale cascade and spectrum follow the same self-similar seasonal cycle \sloppy{(27~days/octave)} but are 40 days out of phase, which suggests that the KE transferred across any mesoscale $\ell$ is primarily deposited at a scale $4\times$ larger. Evolution of   scales larger than $10^3~$km starts being affected by gyre-scale strain, which is spatially inhomogeneous, shaped by domain geometry and wind forcing. 
We found that the upscale mesoscale cascade is dampened substantially at scales of only a few hundred kilometers (Figure~\ref{fig:Volume_Integrated_Pi}). 
This damping may be due to generation of Rossby waves at scales larger than the Rhines scale $\cO(100)~$km, which decorrelates the nonlinear interactions in the meridional direction \cite{VallisMaltrud,Rhines75}. 
However, such decorrelations do not inhibit the upscale cascade from generating larger scales in the east-west direction, for which we find no evidence from our analysis. Moreover, the length-scale at which the mesoscale cascade is arrested decreases poleward (Fig.~\ref{fig:Pi_zonal_means}) unlike the Rhines scale, which does not exhibit significant variations with latitude (see SM Figure~\ref{fig:SUPP:ZonalWithRhines} and \cite{stammer1997global}). These observations reinforce previous studies \cite{sukoriansky2007arrest,khatri2018surface,khatri2019tilted} indicating that Rossby wave generation does not adequately explain the upscale cascade's arrest. 
It is likely that the cascade is attenuated at a few hundred kilometers due to frictional processes that act at these scales in a more isotropic fashion, including eddy damping by wind \cite{Zhai2008,Renault2018,Rai2021} and bottom drag \cite{arbic2009estimates}. 
Determining the relative contribution of these mechanisms to the cascade's ``arrest'' is left to future work. Our findings are consistent with the presence of KE sources and sinks over a wide range of scales in the realistic ocean, unlike in idealized turbulence theories \cite{AlexakisBiferale} where energy is transferred conservatively across scales within the so-called inertial range.

Concerning one motivation in the Introduction on the possibility of energy transfer between mesoscales and the larger coherent circulation leading to intrinsic ocean and climate variability, our demonstration of the existence of such transfer should stimulate further scrutiny into the matter. Indeed, even if the gyre-scales -- mesoscales exchange is a mere fraction of a percent of the $300~$GW mesoscale upscale transfer, it can play a meaningful role in climate dynamics. 
This is because the gyre-scale -- mesoscales KE transfers, while being weaker than those due to self-interactions among the geostrophic mesoscale eddies, probably have a disproportionate effect on the global circulation patterns and climate due to the gyre-scales' coherence in space and persistence in time. 
A rudimentary estimate shows that a net increase of a mere $0.3~$GW  ($0.1\%$ of the $300~$GW) in KE transfer to the gyre-scales of a major circulation pattern, either via amplified upscale transfer or via attenuated downscale transfer, can be momentous. For example, in the case of the ACC's $\approx10~$cm/s gyre-scale ($>10^3~$km) speed \cite{buzzicotti2023coarse}, such a variation would result in a $10\%$ speed increase over 10~years and over a volume of $\approx10^8\,\text{km}^3$ 
assuming energy sinks remain unchanged. This would be a considerable speed-up, $\cO(10)\times$ observed acceleration \cite{shi2021ocean}. Such an estimate is only intended to highlight how minuscule variations in scale-transfer can have a potentially significant impact on gyre-scale circulation, and the importance of advancing rigorous scale-transfer analysis to complement established approaches such as mean-eddy interactions theory and investigations of eddy saturation, standing meanders and bottom topography, among many others\cite{rintoul2018global}. 
Our work lends support to recent results  \cite{Juling2021} showing that the strengthening of mesoscale ocean eddies leads to a strengthening of climate variability. 
Above results also highlight the gyre circulation's departure from linear balance and the potentially important role of effective pressure (normal stress) exerted by the mesoscales, which penetrate the entire water column (Figure~\ref{fig:surface_spectra}), in coupling the ocean surface to the deeper circulation. 
By quantifying the transfer of energy across-scales, we view our work as laying a promising framework for tackling the problem of multiscale coupling within the climate system.

\section*{\MakeUppercase{Materials and Methods}}

\subsection*{NEMO Dataset}
    \label{Methods}
    \label{Methods:dataset}
    The dataset analyzed in this work is publicly available from \hyperlink{https://marine. copernicus.eu/services-portfolio/access-to-products/}{Copernicus Marine Service (CMEMS)}.
    The specific product identifier of the NEMO dataset used here is 
    \\``\texttt{{GLOBAL\_MULTIYEAR\_PHY\_001\_030}}'' \\
    (\hyperlink{https://doi.org/10.48670/moi-00021}{DOI: 10.48670/moi-00021}).
    The ocean reanalysis model has \(\sfrac{1}{12}^\circ\) horizontal grid spacing with 50 vertical levels spanning 5.5~km of depth, with 22 of the vertical levels in the top 100~m.
    Because of the constant land-mask in Antarctica, the NEMO dataset does not extend to \(90^\circ\South\).
    We extend the dataset to the southern pole by extending the land mask (i.e. zero velocity).
    This is done so that filtering kernels intersecting Antarctica are not erroneously truncated.

    \subsubsection*{Surface Results Timeseries}
    \label{Methods:SurfaceTimeseries}
    For results considering only the ocean surface (i.e. \ref{fig:SUPP:Pi_maps_1000km}, \ref{fig:Pi_zonal_means}A-D, and \ref{fig:seasonality_contours}), we analyze the daily-mean ocean velocities at daily resolution for the four years spanning 2015--2018.
    
    \subsubsection*{Depth Results Timeseries}
    \label{Methods:DepthTimeseries}
    Due to the increased computational cost of analyzing full-depth data, when considering depth-dependent results (i.e Figures~\ref{fig:surface_spectra}, \ref{fig:Volume_Integrated_Pi}, and \ref{fig:Pi_zonal_means}E-F) the time-series of full-depth results only includes the first day of each month, as opposed to full daily resolution.
    That is, 01 Jan, 01 Feb, ..., 01 Dec, for the four years spanning 2015--2018.
    
    \subsection*{AVISO Dataset}
    The AVISO dataset analyzed in the Supplementary Information is also available through CMEMS under the product identifier ``\texttt{SEALEVEL\_GLO\_PHY\_L4\_MY\_008\_047}'' (\hyperlink{https://doi.org/10.48670/moi-00145}{DOI: 10.48670/moi-00145}). 
    This dataset provides daily-mean velocities, gridded at \(\sfrac{1}{4}^\circ\) resolution with global coverage.

    \subsection*{Rossby Number}
    \label{Methods:Rossby}
    Figure~\ref{fig:Pi_zonal_means}A-D and the resulting discussion uses the Rossby number (\(\mathrm{Ro}\)), and specifically when \(\mathrm{Ro}=0.1\).
    In this work, the Rossby number is calculated as \(\mathrm{Ro}=U/(f\ell)\), where \(\ell\) is the filter scale, \(f\) is the Coriolis parameter, calculated at each latitude, and \(U=\sqrt{2\rho^{-1}\mathrm{KE^{>\ell}}}\) is the root-mean-squared velocity containing only scales larger than \(\ell\) \cite{Khatri2021}. 
    This approach results in a Rossby number that is a function of both scale and latitude.

    \subsection*{Coarse-graining}
    \label{Methods:CoarseGrain}
    Coarse-graining in simple terms is a convolution between a scalar function being filtered, \(F\), and the filtering kernel \(G\). This can be viewed as a spatially weighted average, performed in a careful manner.
    For a chosen length-scale \(\ell\), given in metres, we define the coarse-grained (i.e. low-pass filtered) scalar function \(\overline{F}_\ell\) as
    \begin{equation}
        \overline{F}_\ell(\vec{x}) = 
        \int_\Omega F(\vec{y})G_{\ell}(\gamma(\vec{x},\vec{y}))dS(\vec{y})
        ,
        \label{eq:CG_definition}
    \end{equation}
    where \(G_{\ell}\) is the filtering kernel, \(dS(\vec{y})\) is the area measure on the sphere, \(\Omega\) is the entire spherical shell, and \(\gamma(\vec{x},\vec{y})\) is geodesic distance between the points \(\vec{x}\) and \(\vec{y}\), 
    \begin{multline}
        \gamma(\vec{x},\vec{y}) = R_{\text{\tiny{E}}}\arccos\!\Big[\sin\phi_x\sin\phi_y
        \\+ \cos\phi_x\cos\phi_y\cos(\lambda_x-\lambda_y)\Big],
        \label{eq:HaversineDistance}
    \end{multline}
    with $R_{\text{\tiny{E}}}=6371~$km for Earth's radius and \(\phi,\lambda\) are the latitude and longitude of \(\vec{x}\) and \(\vec{y}\).    
    
    There are many possible choices for filtering kernel, but desirable properties are \(G(\gamma)\geq0\) for all \(\gamma\) and \(G(\gamma)\to0\) for \(\gamma>\ell/2\). 
    In this work, we use the graded top-hat filter of \cite{Storer2022}, such that the kernel with length-scale \(\ell\) is given by
    \begin{equation}
        G_{\ell}(\gamma) = \frac{A}{2}\left( 1 - \tanh\left( 10\left (\frac{\gamma}{\ell/2}-1 \right ) \right) \right).
        \label{eq:tanh-filter}
    \end{equation}
    In \hyperref[eq:tanh-filter]{eq.~\ref{eq:tanh-filter}}, $A$ is a normalization factor, evaluated numerically, to ensure that $G_\ell$ integrates to unity over the full sphere. 
    The careful mathematical formulation \cite{Aluie2019} and numerical implementation \cite{Storer_FlowSieve} of coarse-graining allows us to preserve flow symmetries at different scales since the operation commutes with derivatives, which enables us to extract meaningful flow diagnostics \cite{Aluie2018}. Ensuring that convolutions and spatial derivatives commute is mathematically non-trivial on the sphere and relies on generalizing the convolution operation \cite{Aluie2019}. The difficulties with commutation are due to curvature, irrespective of any discretization on numerical grids.

     \subsubsection*{Land Treatment}
    \label{sec:LandTreatment}
    Following \cite{Storer2022}, we treat land as zero velocity water for the purpose of coarse-graining, which is consistent with the boundary conditions.
    However, whereas \cite{Storer2022} applied this treatment in the filtering step (i.e. calculating equation~\ref{eq:CG_definition}), in this work it is applied during the Helmholtz projection step, which automatically respects the oceanic flow's boundary conditions. 
    When coarse-graining at a scale $\ell$, the precise boundary between land and ocean becomes blurred at that scale and its precise location becomes less certain. 
    The coarse-grained velocity, $\overline{u}_\ell$, is allowed to be nonzero within a distance $\ell/2$ beyond the continental boundary over land \cite{buzzicotti2023coarse}. 
    Otherwise, requiring $\overline{u}_\ell$ to vanish over land entails deforming the kernel $G_\ell$ in \hyperref[eq:tanh-filter]{eq.~\ref{eq:tanh-filter}}, which breaks the flow symmetries at different scales, \textit{i.e.} coarse-graining would no longer commute with differential operators \cite{Aluie2019}. 
    Forfeiting exact spatial localization in order to gain scale information is theoretically inevitable due to the uncertainty principle \cite{Sadek2018,Storer2022}.
    
    \subsubsection*{Reference Density}
    When calculating \(\mathrm{KE}^{>\ell}\) and \(\Pi_\ell\), we use of the reference density  \(\rho=\rho_0=1025~\mathrm{kg}\cdot\mathrm{m}^{-3}\), as per the Boussinesq approximation used by the NEMO simulation.
    
    \subsubsection*{FlowSieve}
    \label{Methods:FlowSieve}
    The software package used to perform the coarse-graining calculations is FlowSieve \cite{Storer_FlowSieve}, which was developed by the authors. 
    Source code is available at 
    \\\hyperlink{https://github.com/husseinaluie/FlowSieve}{github.com/husseinaluie/FlowSieve}, with documentation at \hyperlink{https://flowsieve.readthedocs.io}{flowsieve.readthedocs.io}.

    \subsection*{Diagnostic Quantities: Filtering Spectrum}
    \label{Methods:Spectrum}
    Following \cite{Sadek2018}, we define the coarse KE per volume as  \(\mathrm{KE}^{>\ell} = 0.5\rho \, \overline{u}_i\overline{u}_i\) (i.e.~the KE of the coarse-grained velocity) and, subsequently, the filtering spectrum as its \(k_{\ell}\)-derivative:
    \begin{equation}
        \OL{E}(k_\ell) = \ddx{k_\ell}\mathrm{KE}^{>\ell} \hspace{1cm}\text{(Filtering Spectrum)},
    \end{equation}
    where \(k_\ell=\ell^{-1}\).
    The filtering spectrum is analogous to the traditional Fourier power spectrum when the latter is valid (c.f. Fig.~4 of \cite{Storer2022}).
    For a positive semi-definite kernel such as the one used here, the steepest resolvable spectral slope is \(-3\) \cite{Sadek2018}.
    However, since we do not measure slopes as steep as \(-3\) (c.f. Figure~\ref{fig:surface_spectra}), this limitation is not a concern for our applications.

    \subsection*{Diagnostic Quantities: KE Scale-Transfer}
    \label{Methods:Cascade}
    KE scale-transfer arising from the non-linear dynamics is obtained from coarse-graining \cite{Germano1992, meneveau1994statistics,eyink1995local,Aluie2009,buzzicotti2018effect} the flow equations as  
    \begin{equation}
        \Pi_\ell := 
        -\frac{\rho}{2}\paren{\overline{u}_{i;j}+\overline{u}_{j;i}}\mathcal{T}^{ij},
        \label{eq:define:Pi}
    \end{equation}
    where repeated indices denote summation, indices after semicolon (;) denote (co-variant) differentiation, and $\mathcal{T}^{ij}$ are (contravariant) components of the sub-scale stress (per unit mass) tensor $\OL{\bu\,\bu}_\ell- \OL{\bu}_\ell\,\OL{\bu}_\ell$ \cite{Aluie2018}. For ocean flows, the term in \hyperref[eq:define:Pi]{eq.~\eqref{eq:define:Pi}} arises from applying the generalized version (see \cite{Aluie2019}) of the convolution in \hyperref[eq:CG_definition]{eq.~\eqref{eq:CG_definition}} to the equation of fluid momentum (per unit mass) on a sphere, 
    \begin{equation}
        \frac{\partial}{\partial t} \overline{\bu}_\ell + \grad\cdot(\OL{\bu}_\ell\,\OL{\bu}_\ell) + \cdots= -\grad\bdot(\OL{\bu\,\bu}_\ell- \OL{\bu}_\ell\,\OL{\bu}_\ell) + \cdots ~,       
        \label{eq:coarseMomentum}
    \end{equation}
    where $\grad\cdot()$ is (co-variant) divergence.
    \hyperref[eq:coarseMomentum]{Eq.~\eqref{eq:coarseMomentum}}, which governs  the flow at scales $>\ell$, is only possible to derive if the convolution commutes with spatial derivatives \cite{Aluie2018,Aluie2019}. Commutation is violated, for example, if filtering is performed by using a spatially varying kernel or by using facile averaging of tensors on the sphere. 
    Taking the inner product of \hyperref[eq:coarseMomentum]{eq.~\eqref{eq:coarseMomentum}} with $\rho\OL\bu$ yields the coarse KE budget, 
    \be
    \partial_t(\rho |\OL{\bu}_\ell|^2/2) + \cdots = -\Pi_\ell + \cdots,
    \ee
    in which $\Pi_\ell$ appears as a sink that is Galilean invariant and scale-local under certain conditions \cite{Aluie2009}. 
    \(\Pi_\ell\) is signed so that positive values indicate energy  transfer from scales larger than \(\ell\) to scales smaller than \(\ell\) \cite{Aluie17}. 
    Where the KE scale-transfer can be reasonably argued to be scale-local, we refer to \(\Pi_\ell\) as the KE `cascade'.
    
    Despite using isotropic kernels for coarse-graining (\hyperref[eq:tanh-filter]{eq.~\ref{eq:tanh-filter}}), $\Pi_\ell$ can still detect anisotropic energy transfer (e.g. due to the Ekman flow) but it, alone, cannot inform us about the direction along which such transfer occurs. To highlight the role of gyre-scale divergence (e.g. due to Ekman transport) in scale-transfer, we can expand \hyperref[eq:define:Pi]{eq.~\eqref{eq:define:Pi}} to obtain 
    \begin{equation}
        \Pi_\ell = -\frac{\rho}{2}\sqp{ \OL{u}_{\lambda;\lambda} \mathcal{T}^{\lambda\lambda} + \OL{u}_{\phi;\phi} \mathcal{T}^{\phi\phi}+\cdots}.
        \label{eq:define:Pi_v2}
    \end{equation}
From \hyperref[eq:define:Pi_v2]{eq.~\eqref{eq:define:Pi_v2}}, a meridionally convergent flow, \textit{i.e.} $\OL{u}_{\phi;\phi} < 0$, yields a positive contribution to $\Pi_\ell$ (\textit{i.e.} downscale KE transfer) since \(\mathcal{T}^{\phi\phi}\) is positive semi-definite and represents a portion of the fine KE, at scales $<\ell$ \cite{Sadek2018,Srinivasan2023}. $\Pi_\ell$ can be written as \(\Pi(\mathbf{u},\mathbf{u},\mathbf{u})\) to highlight its dependence on three velocity modes \cite{Aluie2009}. The first of these modes, \(\Pi(\mathbf{u},\cdot,\cdot)\) contributes to the strain in eq.~\eqref{eq:define:Pi}, while the second and third modes contribute to the subscale stress $\OL{\bu\,\bu}_\ell- \OL{\bu}_\ell\,\OL{\bu}_\ell$ \cite{Aluie2009}.
Given that the lateral velocity \(\mathbf{u}\)  can be decomposed into a laterally non-divergent (T, for toroidal) and laterally divergent (D) components \cite{Aluie2019}, \(\Pi\) can be decomposed exactly into a sum of eight terms: \(\Pi(T,T,T)\), \(\Pi(T,T,D)\), \(\Pi(T,D,T)\), \(\Pi(T,D,D)\), \(\Pi(D,T,T)\), \(\Pi(D,T,D)\), \(\Pi(D,D,T)\), and \(\Pi(D,D,D)\). These terms can represent a different mechanism for KE scale-transfer. For example,
since the non-divergent flow ($T$) is predominantly geostrophic the \(\Pi(T,T,T)\) term expresses transfer due to self-interactions among the geostrophic mesoscale eddies, which generally yield an upscale cascade. By subtracting \(\Pi(T,T,T)\) from \(\Pi\), we are left with the scale-transfer due to interactions involving the divergent flow (e.g. Ekman flow and other unbalanced motions). This is similar, at least in spirit, to the analysis in \cite{Srinivasan2023} at the submesoscales.

    \subsubsection*{Radial/Vertical Velocity}
    In computing both the KE spectrum and scale-transfer, only the zonal and meridional velocity components are considered.
    For hydrostatic flows at scales considered here, the lateral flow makes the overwhelming KE contribution.
    We conducted identical analysis that included radial/vertical velocities diagnosed using flow incompressibility, and found that including the radial velocity \(u_r\) has a negligible impact on both diagnostics across all scales analyzed here.

    \subsubsection*{Ekman Velocity}
    \label{Methods:EkmanVel} Calculations of $\Pi_\ell$ and $\OL{E}(k_\ell)$  relied on the full velocity from the NEMO reanalysis. However, to interpret those results, we sometimes appealed to the Ekman velocity, which is defined as \cite{kang2018extratropical}
    \begin{equation}
    u^E_\lambda = \frac{1}{\rho\cdot{H}^{E}}\frac{f(\phi)}{f(\phi)^2+\epsilon^2}~\tau_\phi \hspace{1.8cm}\text{(zonal)}, 
    \label{eq:EkmanZonalVel}
    \end{equation}
    \begin{equation}
    u^E_\phi = -\frac{1}{\rho\cdot{H}^{E}}\frac{f(\phi)}{f(\phi)^2+\epsilon^2}~\tau_\lambda \hspace{1cm}\text{(meridional)},
    \label{eq:EkmanMeridionalVel}
    \end{equation}
    where \(\rho=\rho_0\) is the reference density, and 
    ${H}^{E}=50~$m is taken to be the gyre-scale Ekman layer depth motivated by Figure~\ref{fig:Pi_zonal_means}F, although taking ${H}^{E}$ to be the seasonally varying mixed layer depth $\mathrm{MLD}$ (Figure~\ref{fig:SUPP:MLD} in SM) yields the same conclusions.
    \(f(\phi)\) is the local Coriolis parameter, \(\epsilon=3.2\times10^{-6}\mathrm{s}^{-1}\approx f(1.25^\circ)\) is the mechanical damping rate \cite{kang2018extratropical}, and \(\tau_\lambda\) is the zonal relative wind stress provided by the NEMO reanalysis data, which follows a bulk formulation
    \begin{equation}
        (\tau_\lambda,\tau_\phi) = \rho_{\mathrm{air}}C_D\absv{\vec{u}_{\mathrm{air}} - \vec{u}_{\mathrm{ocean}}}\paren{\vec{u}_{\mathrm{air}} - \vec{u}_{\mathrm{ocean}}}.
    \label{eq:RelativeWindStress}
    \end{equation}
    The density of air \((\rho_{\mathrm{air}})\), drag coefficient \((C_D)\), and \(10~\mathrm{m}\) air velocity \((\vec{u}_{\mathrm{air}})\) used in the NEMO ocean reanalysis model are obtained from ERA5 atmospheric reanalysis [``\emph{ERA5 hourly data on single levels from 1940 to present}''. Copernicus Climate Change Service (C3S) Climate Data Store (CDS), \hyperlink{https://doi.org/10.24381/cds.adbb2d47}{DOI: 10.24381/cds.adbb2d47}].
    Where necessary, linear interpolation on lat-lon grids was applied to bring the ERA5 data fields onto the same grid as the NEMO ocean data \((\vec{u}_{\mathrm{ocean}})\).

    \subsubsection*{Time-lag and Scale-locality of the Mesoscale Cascade} 
    For the seasonality at each $\ell$, both $-\Pi(\ell,t)$ and $\OL{E}(\ell,t)$ have a period of 365~days but $\OL{E}(\ell,t)$ is phase-shifted 41~days after $-\Pi(\ell,t)$. 
    Therefore, the KE spectrum tendency, $\frac{d}{dt}\OL{E}(\ell,t)$, is phase-shifted $-41+365/4=50~$days \emph{before} $-\Pi(\ell,t)$. 
    Noting that the cycle for each of $-\Pi(2\ell,t)$ and $\frac{d}{dt}\OL{E}(2\ell,t)$ at scale $2\ell$  is phase-shifted 27~days later relative to that at scale $\ell$, we have that the upscale cascade, $-\Pi(\ell,t)$, is in-phase with $\frac{d}{dt}\OL{E}(3.6\ell,t)$ at length-scale $3.6\ell\approx4\ell$.

    \subsubsection*{Horizontal, Vertical, and Temporal Averages and Integrals}
    The diagnostics variables are computed at all points in space and time for the entire dataset of consideration. That is, \(\Pi_\ell=\Pi_\ell(t,z,\phi,\lambda)\), for time \(t\), depth \(z\), latitude \(\phi\), and longitude \(\lambda\).
    \begin{itemize}
    \item{
        Horizontal averages (such as Figures~\ref{fig:surface_spectra}) are spatial integrals normalized by the spatial area, with appropriate weighting by the cell area:
        \begin{equation}
        \text{\small{HorizAvg}}(F)(t,z,\Omega) = \frac{\int_{(\phi,\lambda)\in\Omega}F(t,z,\phi,\lambda)dA}{\absv{\Omega}},
        \label{eq:horizontal_average}
        \end{equation}
        where \(\Omega\) is the spatial region of interest and 
        \[\absv{\Omega}=\int_{(\phi,\lambda)\in\Omega}\mathrm{IsWater}(\phi,\lambda,z)dA\] 
        is the \emph{water} area (i.e. \(\mathrm{IsWater}(\phi,\lambda,z)=1\) if \((\phi,\lambda,z)\) is a water cell and \(0\) otherwise.
        Horizontal averages are then functions of time, depth, and the choice of region.
        Horizontal integrals remove the normalization factor \(\absv{\Omega}^{-1}\).
    }
    \item{ 
        Zonal averages (such as Figure~\ref{fig:Pi_zonal_means}A-F), are computed along lines of constant latitude.
        As with horizontal averages, zonal averages are normalized by \emph{water} area at each latitude.
    }
    \item{
        Depth integrals account for the depth-varying vertical thickness of cell grids by treating z-levels as cell bottoms, and extending the top cell (depth of \(\sim0.5~\)m) to the surface.
        Vertical thicknesses vary monotonically from \(\sim0.5~\)m to \(\sim450~\)m.
    }
    \item{ 
        Time averages (both means and medians) are computed in the standard way, since we have uniform time sampling.
        In the case of seasonal averages (e.g. Figure~\ref{fig:Volume_Integrated_Pi}), the time-series is partitioned based on the month of the year, with the mean/median of each partition computed separately.
    }
    \end{itemize} 
    
    \subsection*{Helmholtz Decomposition}
    \label{Methods:Helmholtz}
    Unlike the analysis in \cite{Storer2022} using geostrophic velocity, the results presented here use the full horizontal model velocity, which contains both rotational and horizontally divergent components. This generality renders coarse-graining the velocity field in a manner that commutes with spatial derivatives more complicated, involving the so-called Edmonds transformation \cite{Aluie2019}.
    A solution we use here is to first perform a Helmholtz decomposition of the velocity field and obtain coarse velocities from the coarse-grained Helmholtz scalars, which is equivalent to performing the Edmonds transformation \cite{Aluie2019}.
    Specifically, if
    \renewcommand{\arraystretch}{1.4}
    \begin{equation}
        \vec{u} = \sqp{ \begin{array}{c}u_\lambda\\u_\phi\end{array} } 
        = 
        \sqp{ \begin{array}{cc}
            -\ppx{\phi} & \sec(\phi)\ppx{\lambda} \\ 
            \sec(\phi)\ppx{\lambda} & \ppx{\phi}
            \end{array}}
        \sqp{ \begin{array}{c}\Psi\\\Phi\end{array}}
        \label{eq:HelmholtzRelation}
    \end{equation}
    \renewcommand{\arraystretch}{1.}
    where \(\lambda,\phi\) are the longitude and latitude, \(u_\lambda,u_\phi\) are the zonal and meridional velocities, and \(\Psi,\Phi\) are the Helmholtz scalars, then
    \renewcommand{\arraystretch}{1.4}
    \begin{equation}
        \overline{\vec{u}}
        = 
        \sqp{ \begin{array}{cc}
            -\ppx{\phi} & \sec(\phi)\ppx{\lambda} \\ 
            \sec(\phi)\ppx{\lambda} & \ppx{\phi}
            \end{array}}
        \sqp{ \begin{array}{c}\overline{\Psi}\\\overline{\Phi}\end{array}}
    \end{equation}
    \renewcommand{\arraystretch}{1.}
    where \(\overline{\Psi},\overline{\Phi}\) are computed by coarse-graining each field as a scalar \cite{Aluie2019,Storer_FlowSieve}.
    Computational details of the Helmholtz decomposition can be found in the Supplemental Information (SM).

\clearpage

\bibliography{scibib}

\begin{thebibliography}{10}

\bibitem{winton2003climatic}
M.~Winton, {\it Journal of Climate\/} {\bf 16}, 2875 (2003).

\bibitem{talley2011descriptive}
L.~D. Talley, {\it {Descriptive physical oceanography: an introduction}\/}
  (Academic Press, 2011).

\bibitem{Storer2022}
B.~A. Storer, M.~Buzzicotti, H.~Khatri, S.~M. Griffies, H.~Aluie, {\it Nature
  Communications\/} {\bf 13}, 5314 (2022).

\bibitem{Chelton2011}
D.~B. Chelton, M.~G. Schlax, R.~M. Samelson, {\it Progress in Oceanography\/}
  {\bf 91}, 167 (2011).

\bibitem{Arbic2014}
B.~K. Arbic, {\it et~al.\/}, {\it Journal of Physical Oceanography\/} {\bf 44},
  2050 (2014).

\bibitem{serazin2018inverse}
G.~S{\'{e}}razin, {\it et~al.\/}, {\it Journal of Physical Oceanography\/} {\bf
  48}, 1385 (2018).

\bibitem{arzel2020contributions}
O.~Arzel, T.~Huck, {\it Journal of Climate\/} {\bf 33}, 2351 (2020).

\bibitem{gehlen2020quantification2}
M.~Gehlen, S.~Berthet, R.~S{\'e}f{\'e}rian, C.~Eth{\'e}, T.~Penduff, {\it
  Geophysical Research Letters\/} {\bf 47}, e2020GL088304 (2020).

\bibitem{Juling2021}
A.~J{\"{u}}ling, A.~von~der Heydt, H.~A. Dijkstra, {\it Ocean Science\/} {\bf
  17}, 1251 (2021).

\bibitem{constantinou2021intrinsic}
N.~C. Constantinou, A.~M. Hogg, {\it Journal of Climate\/} {\bf 34}, 6175
  (2021).

\bibitem{hochet2022energy}
A.~Hochet, T.~Huck, O.~Arzel, F.~S{\'e}vellec, A.~C.~d. Verdi{\`e}re, {\it
  Journal of Climate\/} {\bf 35}, 1157 (2022).

\bibitem{Vallis2006}
G.~K. Vallis, {\it {Atmospheric and Oceanic Fluid Dynamics}\/} (Cambridge
  University Press, Cambridge, U.K., 2006).

\bibitem{richardson1922weather}
L.~F. Richardson, {\it {Weather prediction by numerical process}\/} (University
  Press, 1922).

\bibitem{Kolmogorov41a}
A.~N. Kolmogorov, {\it Doklady Akademii Nauk SSSR\/} {\bf 30}, 9 (1941).

\bibitem{Onsager49}
L.~Onsager, {\it Il Nuovo Cimento\/} {\bf 6}, 279 (1949).

\bibitem{eyink2006onsager}
G.~L. Eyink, K.~R. Sreenivasan, {\it Reviews of modern physics\/} {\bf 78}, 87
  (2006).

\bibitem{AlexakisBiferale}
A.~Alexakis, L.~Biferale, {\it Physics Reports\/} {\bf 767-769}, 1 (2018).

\bibitem{zhou2021turbulence}
Y.~Zhou, {\it Physics Reports\/} {\bf 935}, 1 (2021).

\bibitem{fjortoft1953changes}
R.~Fj{\o}rtoft, {\it Tellus\/} {\bf 5}, 225 (1953).

\bibitem{Kraichnan1967}
R.~H. Kraichnan, {\it Physics of Fluids (1958-1988)\/} {\bf 10}, 1417 (1967).

\bibitem{charney1971geostrophic}
J.~G. Charney, {\it Journal of the Atmospheric Sciences\/} {\bf 28}, 1087
  (1971).

\bibitem{Ferrari2009}
R.~Ferrari, C.~Wunsch, {\it Annual Review of Fluid Mechanics\/} {\bf 41}, 253
  (2009).

\bibitem{Wunsch1998}
C.~Wunsch, {\it Journal of Physical Oceanography\/} {\bf 28}, 2332 (1998).

\bibitem{Hughes2008}
C.~W. Hughes, C.~Wilson, {\it Journal of Geophysical Research\/} {\bf 113},
  C02016 (2008).

\bibitem{arbic2009estimates}
B.~K. Arbic, {\it et~al.\/}, {\it Journal of Geophysical Research\/} {\bf 114},
  C02024 (2009).

\bibitem{Nikurashin2011}
M.~Nikurashin, R.~Ferrari, {\it Geophysical Research Letters\/} {\bf 38}, n/a
  (2011).

\bibitem{Scott2005}
R.~B. Scott, F.~Wang, {\it Journal of Physical Oceanography\/} {\bf 35}, 1650
  (2005).

\bibitem{Arbic2013}
B.~K. Arbic, K.~L. Polzin, R.~B. Scott, J.~G. Richman, J.~F. Shriver, {\it
  Journal of Physical Oceanography\/} {\bf 43}, 283 (2013).

\bibitem{qiu2014seasonal}
B.~Qiu, S.~Chen, P.~Klein, H.~Sasaki, Y.~Sasai, {\it Journal of Physical
  Oceanography\/} {\bf 44}, 3079 (2014).

\bibitem{sasaki2014impact}
H.~Sasaki, P.~Klein, B.~Qiu, Y.~Sasai, {\it Nature Communications\/} {\bf 5},
  5636 (2014).

\bibitem{khatri2018surface}
H.~Khatri, J.~Sukhatme, A.~Kumar, M.~K. Verma, {\it Journal of Geophysical
  Research: Oceans\/} {\bf 123}, 3875 (2018).

\bibitem{Balwada2022}
D.~Balwada, J.-H. Xie, R.~Marino, F.~Feraco, {\it Science Advances\/} {\bf 8},
  eabq2566 (2022).

\bibitem{Aluie2018}
H.~Aluie, M.~Hecht, G.~K. Vallis, {\it Journal of Physical Oceanography\/} {\bf
  48}, 225 (2018).

\bibitem{Sadek2018}
M.~Sadek, H.~Aluie, {\it Physical Review Fluids\/} {\bf 3}, 1 (2018).

\bibitem{Xu2012}
Y.~Xu, L.~L. Fu, {\it Journal of Physical Oceanography\/} {\bf 42}, 2229
  (2012).

\bibitem{Uchida2017}
T.~Uchida, R.~Abernathey, S.~Smith, {\it Ocean Modelling\/} {\bf 118}, 41
  (2017).

\bibitem{Khatri2021}
H.~Khatri, S.~M. Griffies, T.~Uchida, H.~Wang, D.~Menemenlis, {\it Geophysical
  Research Letters\/} {\bf 48}, 1 (2021).

\bibitem{Ajayi2021}
A.~Ajayi, {\it et~al.\/}, {\it Journal of Advances in Modeling Earth Systems\/}
  {\bf 13} (2021).

\bibitem{Callies2013}
J.~Callies, R.~Ferrari, {\it Journal of Physical Oceanography\/} {\bf 43}, 2456
  (2013).

\bibitem{rintoul2018global}
S.~R. Rintoul, {\it Nature\/} {\bf 558}, 209 (2018).

\bibitem{lacasce2020baroclinic}
J.~H. LaCasce, S.~Groeskamp, {\it Journal of Physical Oceanography\/} {\bf 50},
  2835 (2020).

\bibitem{buzzicotti2023coarse}
M.~Buzzicotti, B.~Storer, H.~Khatri, S.~Griffies, H.~Aluie, {\it Journal of
  Advances in Modeling Earth Systems\/} {\bf 15}, e2023MS003693 (2023).

\bibitem{tulloch2011scales}
R.~Tulloch, J.~Marshall, C.~Hill, K.~S. Smith, {\it Journal of Physical
  Oceanography\/} {\bf 41}, 1057 (2011).

\bibitem{Srinivasan2023}
K.~Srinivasan, R.~Barkan, J.~C. McWilliams, {\it Journal of Physical
  Oceanography\/} {\bf 53}, 287 (2023).

\bibitem{kraichnan1974kolmogorov}
R.~H. Kraichnan, {\it Journal of Fluid Mechanics\/} {\bf 62}, 305 (1974).

\bibitem{Aluie11}
H.~{Aluie}, {\it Phys. Rev. Lett.\/} {\bf 106}, 174502 (2011).

\bibitem{Aluie13}
H.~Aluie, {\it Physica D: Nonlinear Phenomena\/} {\bf 247}, 54 (2013).

\bibitem{buzzicotti2016phase}
M.~Buzzicotti, B.~P. Murray, L.~Biferale, M.~D. Bustamante, {\it The European
  Physical Journal E\/} {\bf 39}, 1 (2016).

\bibitem{schubert2020submesoscale}
R.~Schubert, J.~Gula, R.~J. Greatbatch, B.~Baschek, A.~Biastoch, {\it Journal
  of Physical Oceanography\/} {\bf 50}, 2573 (2020).

\bibitem{Salmon80}
R.~Salmon, {\it Geophysical {\&} Astrophysical Fluid Dynamics\/} {\bf 15}, 167
  (1980).

\bibitem{Menesguen2019}
C.~M{\'{e}}nesguen, {\it et~al.\/}, {\it Earth and Space Science\/} {\bf 6},
  370 (2019).

\bibitem{Hua2008}
B.~L. Hua, {\it et~al.\/}, {\it {Destabilization of mixed Rossby gravity waves
  and the formation of equatorial zonal jets}\/}, vol. 610 (2008).

\bibitem{CliftPlumb2008monsoon}
P.~D. Clift, R.~A. Plumb, {\it The Asian monsoon: causes, history and
  effects\/} (Cambridge University Press Cambridge, London, 2008), first edn.

\bibitem{schneider2014migrations}
T.~Schneider, T.~Bischoff, G.~H. Haug, {\it Nature\/} {\bf 513}, 45 (2014).

\bibitem{kang2018extratropical}
S.~M. Kang, Y.~Shin, S.-P. Xie, {\it Npj Climate and Atmospheric Science\/}
  {\bf 1}, 20172 (2018).

\bibitem{mitchell1992annual}
T.~P. Mitchell, J.~M. Wallace, {\it Journal of Climate\/} {\bf 5}, 1140 (1992).

\bibitem{philander1996itcz}
S.~Philander, {\it et~al.\/}, {\it Journal of climate\/} {\bf 9}, 2958 (1996).

\bibitem{callies2015seasonality}
J.~Callies, R.~Ferrari, J.~M. Klymak, J.~Gula, {\it Nature communications\/}
  {\bf 6}, 6862 (2015).

\bibitem{chen2006physical}
S.~Chen, {\it et~al.\/}, {\it Physical review letters\/} {\bf 96}, 084502
  (2006).

\bibitem{loose2023diagnosing}
N.~Loose, S.~Bachman, I.~Grooms, M.~Jansen, {\it Journal of Physical
  Oceanography\/} {\bf 53}, 157 (2023).

\bibitem{Zhai2008}
X.~Zhai, R.~J. Greatbatch, J.~D. Kohlmann, {\it Geophysical Research Letters\/}
  {\bf 35}, 1 (2008).

\bibitem{Renault2018}
L.~Renault, J.~C. McWilliams, J.~Gula, {\it Scientific Reports\/} {\bf 8}, 1
  (2018).

\bibitem{Rai2021}
S.~Rai, M.~Hecht, M.~Maltrud, H.~Aluie, {\it Science Advances\/} {\bf 7}, 1
  (2021).

\bibitem{VonStorch2012}
J.~S. Von~Storch, {\it et~al.\/}, {\it Journal of Physical Oceanography\/} {\bf
  42}, 2185 (2012).

\bibitem{scott2009update}
R.~B. Scott, Y.~Xu, {\it Deep Sea Research Part I: Oceanographic Research
  Papers\/} {\bf 56}, 295 (2009).

\bibitem{Aluie2019}
H.~Aluie, {\it GEM - International Journal on Geomathematics\/} {\bf 10}, 9
  (2019).

\bibitem{Storer_FlowSieve}
B.~A. Storer, H.~Aluie, {\it Journal of Open Source Software\/} {\bf 8}, 4277
  (2023).

\bibitem{VallisMaltrud}
G.~K. Vallis, M.~E. Maltrud, {\it Journal of Physical Oceanography\/} {\bf 23},
  1346 (1993).

\bibitem{Rhines75}
P.~B. Rhines, {\it Journal of Fluid Mechanics\/} {\bf 69}, 417–443 (1975).

\bibitem{stammer1997global}
D.~Stammer, {\it Journal of Physical Oceanography\/} {\bf 27}, 1743 (1997).

\bibitem{sukoriansky2007arrest}
S.~Sukoriansky, N.~Dikovskaya, B.~Galperin, {\it Journal of the Atmospheric
  Sciences\/} {\bf 64}, 3312 (2007).

\bibitem{khatri2019tilted}
H.~Khatri, P.~Berloff, {\it Journal of Fluid Mechanics\/} {\bf 876}, 939
  (2019).

\bibitem{shi2021ocean}
J.-R. Shi, L.~D. Talley, S.-P. Xie, Q.~Peng, W.~Liu, {\it Nature Climate
  Change\/} {\bf 11}, 1090 (2021).

\bibitem{Germano1992}
M.~Germano, {\it Journal of Fluid Mechanics\/} {\bf 238}, 325 (1992).

\bibitem{meneveau1994statistics}
C.~Meneveau, {\it Physics of Fluids\/} {\bf 6}, 815 (1994).

\bibitem{eyink1995local}
G.~L. Eyink, {\it Journal of Statistical Physics\/} {\bf 78}, 335 (1995).

\bibitem{Aluie2009}
H.~Aluie, G.~L. Eyink, {\it Physics of Fluids\/} {\bf 21}, 1 (2009).

\bibitem{buzzicotti2018effect}
M.~Buzzicotti, {\it et~al.\/}, {\it Journal of Turbulence\/} {\bf 19}, 167
  (2018).

\bibitem{Aluie17}
H.~Aluie, {\it New Journal of Physics\/} {\bf 19}, 025008 (2017).

\bibitem{Bochkanov}
S.~Bochkanov, {ALGLIB} (2022).

\end{thebibliography}
\bibliographystyle{plain}

\paragraph{Acknowledgments} 
We thank Houssam Yassin and two anonymous reviewers for their thoughtful feedback.
We used the \href{https://pilestone.com/pages/color-blindness-simulator-1}{Pilestone color blind vision simulator} to test our presented figures.

\paragraph{Funding} 
    This research was funded by US NASA grant 80NSSC18K0772 and NSF grant OCE-2123496. 
    HA was also supported by US DOE grants DE-SC0014318, DE-SC0020229, DE-SC0019329, NSF grants PHY-2020249, PHY-2206380, and US NNSA grants DE-NA0003856, DE-NA0003914, DE-NA0004134. HK acknowledges the support from UK NERC grant NE/T013494/1.  
    SMG acknowledges support from the National Oceanic and Atmospheric Administration Geophysical Fluid Dynamics Laboratory. 
    MB acknowledges support from the European Research Council (ERC) under the European Union’s Horizon 2020 research and innovation programme (Grant Agreement No. 882340). 
    Computing time was provided by NERSC under Contract No. DE-AC02-05CH11231, NASA's HEC Program through NCCS at Goddard Space Flight Center, and the Texas Advanced Computing Center (TACC) at The University of Texas at Austin, under ACCESS allocation grant EES220052.
    The statements, findings, conclusions, and recommendations are those of the author(s) and do not necessarily reflect the views of the National Oceanic and Atmospheric Administration, or the U.S. Department of Commerce.

\paragraph{Competing Interests} The authors declare they have no competing interest.

\paragraph{Data and Materials Availability}
The FlowSieve package \cite{Storer_FlowSieve} developed by the authors and used to perform the coarse-graining calculations is publicly available at \hyperlink{https://github.com/husseinaluie/FlowSieve}{github.com/husseinaluie/FlowSieve}.
We will also share publicly the post-processed data that we used for these results, as well as Jupyter notebooks that reproduce the figures during the review.

\onecolumn
\appendix
 
\makeatletter 
\renewcommand{\thefigure}{S\@arabic\c@figure}
\makeatother
\setcounter{figure}{0} 
\renewcommand{\theequation}{S-\arabic{equation}}
\setcounter{equation}{0}  
\renewcommand{\thepage}{{\it Supplementary Material -- \arabic{page}}} \setcounter{page}{1}

\section*{\MakeUppercase{Supplementary Material}}

\setcounter{equation}{0}  
\renewcommand{\theequation}{S-\arabic{equation}}

\setcounter{page}{1}  
\renewcommand{\thepage}{SM-\arabic{page}}

\subsection*{\MakeUppercase{Power Spectra at Various Depth}}

\begin{figure}[tbhp]
    \centering
    \includegraphics[scale=1.35]{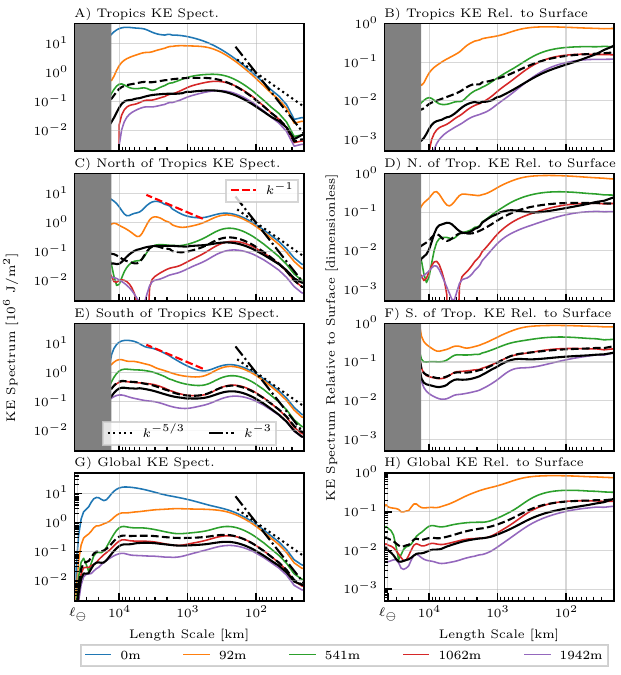}
    \caption{ 
        \textbf{KE Spectra at Selected Depths}
        [A,C,E,G] Similar structure to Figure~\ref{fig:surface_spectra}, but with spectra plotted only for selected depths [see in-set legend].
        [B,D,F,H] are again similarly structured, but now showing the KE spectrum divided by the surface spectrum.
        [Thick solid black lines] show spectra of the depth-averaged flow, while [thick dashed black lines] show the depth-averaged spectra. 
    }
    \label{fig:spectra:SUPP}
\end{figure}

Figure~\ref{fig:spectra:SUPP} is supplemental to Figure~\ref{fig:surface_spectra}, and shows the power spectra of selected depths, along with the spectrum of the depth-average flow (solid black lines) and the depth-average of the power spectra (dashed black lines).

\subsection*{\MakeUppercase{Helmholtz Computational Framework}}
    The Helmholtz system to be solved is given by \hyperref[eq:HelmholtzProject-simple]{eq.~\ref{eq:HelmholtzProject-simple}}. 
    Nominally, those two equations alone would be sufficient to determine the Helmholtz scalars.
    \renewcommand{\arraystretch}{1.4}
    \begin{equation}
        \sqp{\begin{array}{cc}
        \nabla^2  & 0 \\
        0         & \nabla^2                  
        \end{array}}
        \sqp{\begin{array}{c}\Psi\\\Phi\end{array}} 
        = 
        \sqp{\begin{array}{c}\zeta\\\delta\end{array}}
        \label{eq:HelmholtzProject-simple}
    \end{equation}
    \renewcommand{\arraystretch}{1.}
    However, because the second derivatives both share even symmetry, spurious grid-scale noise can arise.
    To resolve this, two additional rows are added, yielding \hyperref[eq:HelmholtzProject]{eq.~\ref{eq:HelmholtzProject}}.
    The first two rows of \hyperref[eq:HelmholtzProject]{eq.~\ref{eq:HelmholtzProject}} define the relationship between the velocities, \(u_\lambda,u_\phi\), and the Helmholtz scalars, \(\Psi,\Phi\) (i.e. \hyperref[eq:HelmholtzRelation]{eq.~\ref{eq:HelmholtzRelation}}), and the second two impose that the vorticity \(\zeta\) and divergence \(\delta\) be wholly described by \(\Psi\) and \(\Phi\) respectively.
    Using both the upper and lower halves of the least-squares problem reduces computationally spurious noise by including both first derivatives, which have odd-symmetry stencils, and second order derivatives, which have even-symmetry stencils. 
    The scaling factor \(\alpha\) allows tuning between the upper and lower halves of the projection operator, while the Laplacian is given by 
    \[ \nabla^2 := (\cos\phi)^{-2}\partial^2_{\lambda\lambda} + \partial^2_{\phi\phi} - \tan(\phi)\partial_{\phi}. \]
    \renewcommand{\arraystretch}{1.4}
    \begin{equation}
        \sqp{\begin{array}{cc}
        -\partial_{\phi}                & \sec(\phi) \partial_{\lambda} \\
        \sec(\phi) \partial_{\lambda}   & \partial_{\phi}               \\
        \alpha\nabla^2                  & 0                             \\
        0                               & \alpha\nabla^2                  
        \end{array}}
        \sqp{\begin{array}{c}\Psi\\\Phi\end{array}} 
        = 
        \sqp{\begin{array}{c}u_\lambda\\u_\phi\\\alpha\zeta\\\alpha\delta\end{array}}
        \label{eq:HelmholtzProject}
    \end{equation}
    \renewcommand{\arraystretch}{1.}
    The Helmholtz scalars are computed using an iterative least-squares solver.
    Specifically, \hyperref[eq:HelmholtzProject]{eq.~\ref{eq:HelmholtzProject}} is converted into a sparse matrix problem by replacing the differential operators with matrix operators using finite difference approximations.
    The results presented in this work used a fourth-order finite difference scheme.
    The solve step is performed using the sparse least-squares solver provided in ALGLIB \cite{Bochkanov}: \texttt{linlsqrsolvesparse}, a matrix-free iterative solver.
    Convergence of the iterative solver is improved by first downsampling the velocities \(u_\lambda,u_\phi\) onto coarser grids, solving for the Helmholtz scalars on the coarse grid, and providing the coarse solution as an initial guess for the higher resolution solve.
    
\subsubsection*{\MakeUppercase{Boundary Conditions}}
    Our treatment of land cells as zero-velocity water cells automatically imposes the boundary conditions required for solving \hyperref[eq:HelmholtzProject-simple]{eq.~\ref{eq:HelmholtzProject-simple}} (or \hyperref[eq:HelmholtzProject]{eq.~\ref{eq:HelmholtzProject}}), which we do over the entire spherical domain.
    
\subsubsection*{\MakeUppercase{Building the Least-Squares Matrices}}
    The entries in \hyperref[eq:HelmholtzProject]{eq.~S-2} are block matrices.
    That is, each entry itself represents a matrix constructed using the standard Kronecker product method, outlined here for the purpose of completeness.
    Suppose the grid has \(N_\phi\), \(N_\lambda\) points in latitude and longitude.
    Then let \(I^\phi\), \(I^\lambda\) be the \(N_\phi\times N_\phi\) and \(N_\lambda\times N_\lambda\) identity matrices.
    Let \(D^\phi\), \(D^\lambda\) be the finite difference first-derivative matrices on the \(\phi\) and \(\lambda\) grids (of size \(N_\phi\times N_\phi\) and \(N_\lambda\times N_\lambda\), respectively).
    The entries of \hyperref[eq:HelmholtzProject]{eq.~S-2} are then built using Kronecker products: \(\partial_\phi=\mathrm{Kron}(D^\phi,I^\lambda)\), \(\partial_\lambda=\mathrm{Kron}(I^\phi,D^\lambda)\), etc., so that each sub-array is of size \(N_\phi N_\lambda\times N_\phi N_\lambda\).
    The final least-squares problem is a matrix of size \(4N_\phi N_\lambda\times 2N_\phi N_\lambda\), with the solution array a vector of length \(2N_\phi N_\lambda\).
    Since we use fourth-order finite difference derivatives, the least squares matrix is very sparse.

\subsection*{\MakeUppercase{Mixed Layer Depth}}
\begin{figure}[tbhp]
    \centering
    \includegraphics[scale=1]{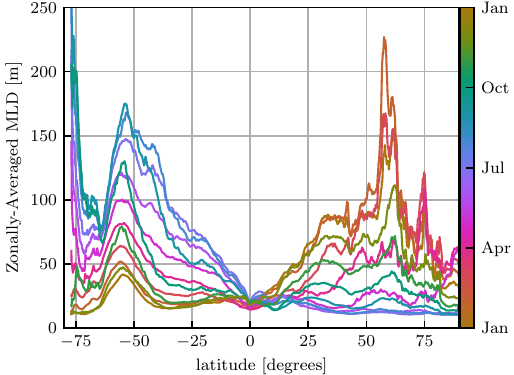}
    \caption{ 
        \textbf{Mixed Layer Depth} 
        The monthly-mean zonally-averaged mixed layer depth (MLD).  It is provided in the NEMO dataset and corresponds to 2018.
    }
    \label{fig:SUPP:MLD}
\end{figure}

Figure~\ref{fig:SUPP:MLD} presents the monthly-mean Mixed Layer Depth (MLD) as a function of latitude. 
The lines illustrate how the MLD can increase 3--4 times during the local winter compared to the local summer.

\subsection*{\MakeUppercase{Annotated Zonally Averaged \(\Pi\)}}
\begin{figure}[tbhp]
    \centering
    \includegraphics[width=7.4cm]{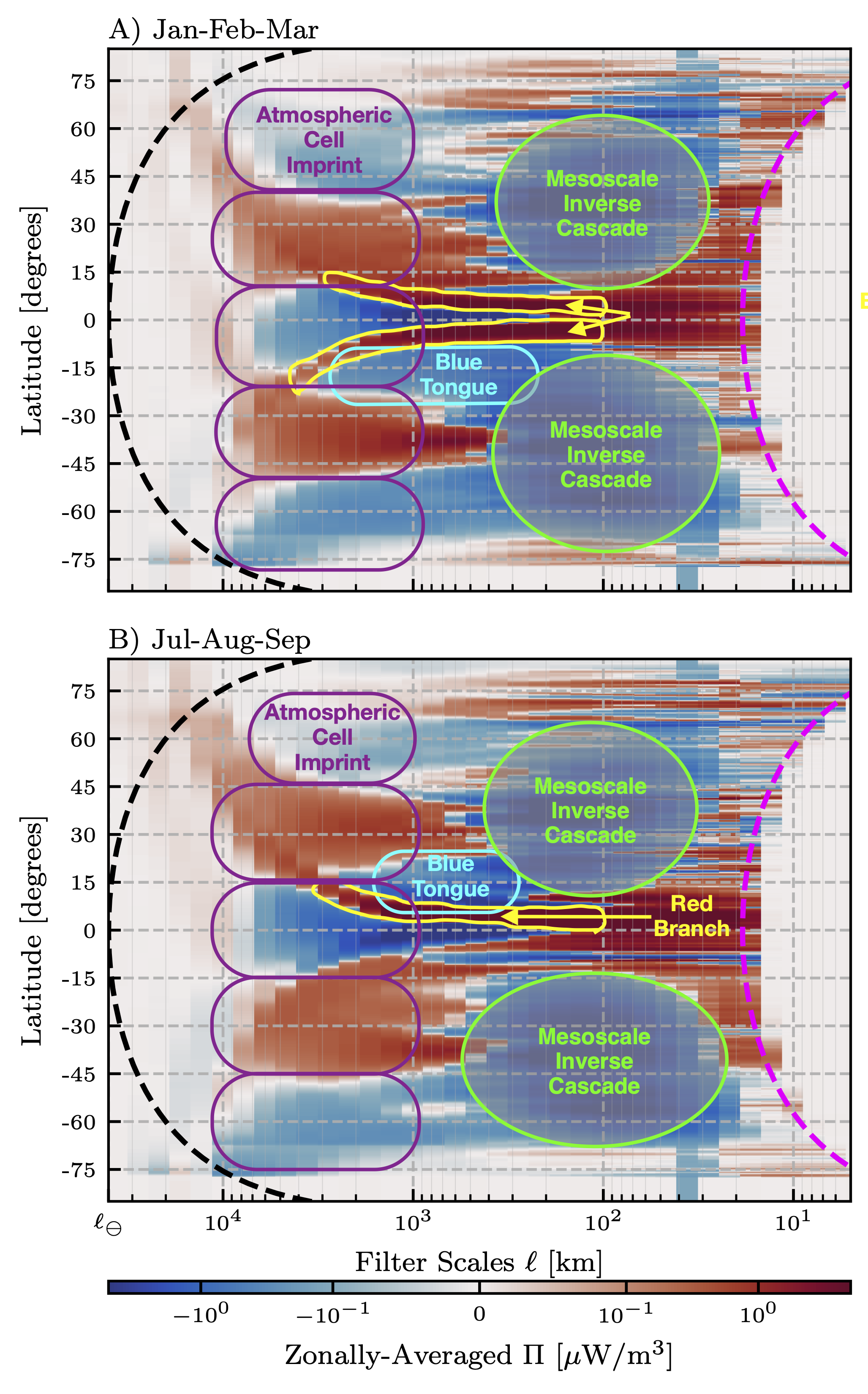}
    \caption{ 
        \textbf{Annotated version of zonal mean \(\Pi\)}
        Reproduction of Figure~\ref{fig:Pi_zonal_means}C-D with additional annotations overlain.
        The annotations are:
        [green circles] show the mesoscale inverse cascade,
        [blue rounded boxes] show the ``blue tongue,''
        [yellow oblong shapes] show that ``red branches'', and
        [purple rounded boxes] show the Ekman pattern from the atmospheric cells.
    }
    \label{fig:SUPP:zonal_Pi:annotated}
\end{figure}

Figure~\ref{fig:SUPP:zonal_Pi:annotated} provides an annotated version of Figure~\ref{fig:Pi_zonal_means}C-D, with annotations showing the \textbf{[green circles]} show the mesoscale inverse cascade,\textbf{[blue rounded boxes]} show the ``blue tongue'', \textbf{[yellow oblong shapes]} show that ``red branches'', and \textbf{[purple rounded boxes]} show the Ekman pattern from the atmospheric cells.
The annotations are purely qualitative and for the purpose of illustration.

\subsection*{\MakeUppercase{Rhines Scale}}
\begin{figure}
    \centering
    \includegraphics{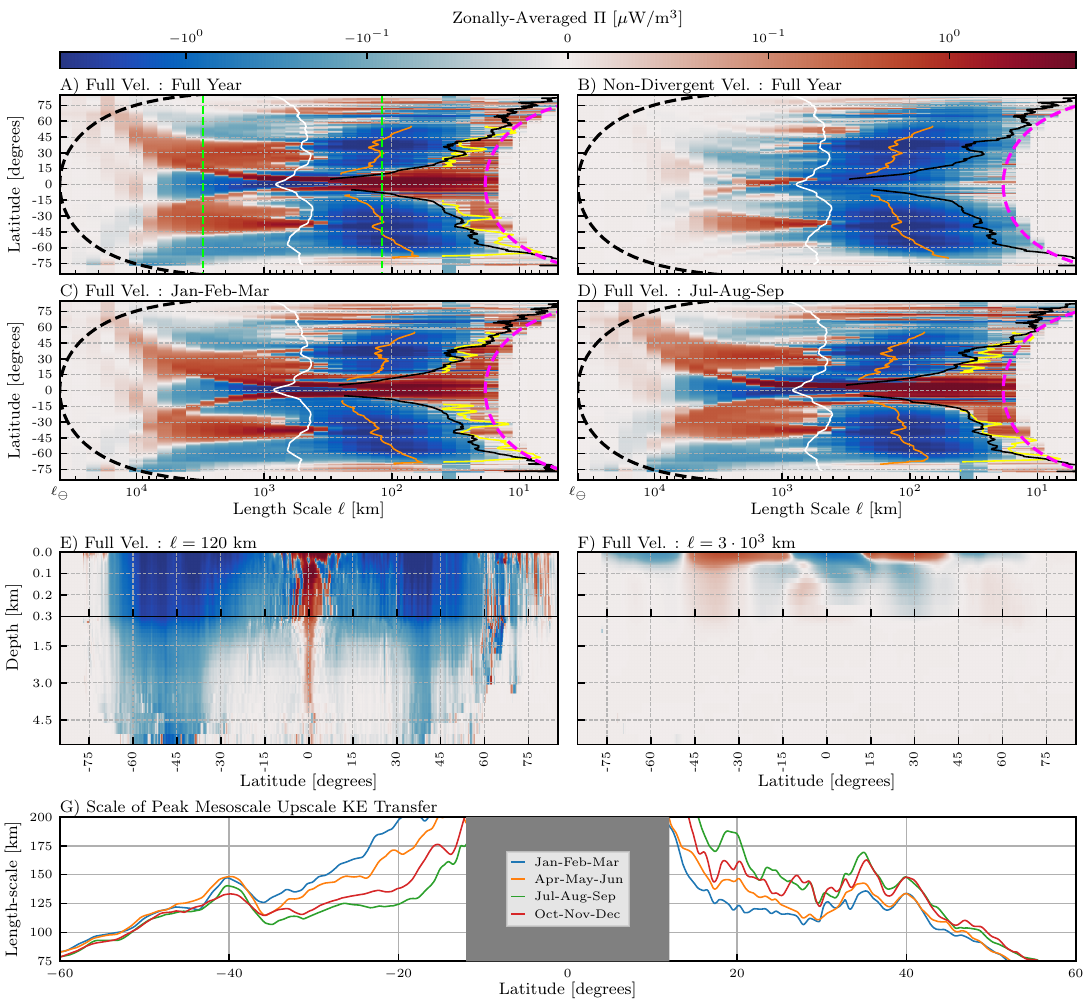}
    \caption{ A reproduction of Figure~\ref{fig:Pi_zonal_means}, with panels [A-D] now included a solid  white contour line that shows \(\ell_{\mathrm{Rhines}}\).}
    \label{fig:SUPP:ZonalWithRhines}
\end{figure}

With coarse-graining, we can define the Rhines scale \cite{VallisMaltrud,Rhines75} as a function of latitude (\(\phi\)) as the solution to the following implicit equation,  
\be
\ell_\mathrm{Rhines}(\phi) = 2\pi\sqrt{u_{\mathrm{rms}}(\ell_\mathrm{Rhines},\phi)/\beta(\phi)},
\lb{eq:RhinesScale}\ee
where \(u_{\mathrm{rms}}(\ell,\phi)=\sqrt{2\rho^{-1}\mathrm{KE^{>\ell}}(\phi)}\) is the rms-velocity of all scales larger than \(\ell\) at latitude  $\phi$. To solve eq.~\eqref{eq:RhinesScale}, we evaluate the right-hand-side (RHS) for each $\phi$ over the entire range of scales $\ell$ and find where the RHS equals $\ell$.
From the NEMO data, we find that  \(\ell_\mathrm{Rhines}(\phi)\approx500\pm100~\)km, without any obvious dependence on latitude $\phi$. This result is consistent with previous estimates of the Rhines scale to be $\cO(100)~$km without a clear variation with latitude (Fig.~25 in \cite{stammer1997global}).
However, Figure~\ref{fig:Pi_zonal_means} shows that the length-scale at which the mesoscale upscale cascade is arrested decreases poleward, which is not reflected in the Rhines scale. This result suggests that the generation of Rossby waves, also known as the $\beta$-effect \cite{VallisMaltrud,Rhines75}, is probably not the main mechanism by which the mesoscale cascade is arrested. 
Figure~\ref{fig:SUPP:ZonalWithRhines} reproduces Figure~\ref{fig:Pi_zonal_means} with an additional line showing the Rhines scale as a function of latitude. The Rhines scale is found to be mostly in the interval \([400~\mathrm{km},600~\mathrm{km}]\) and, unlike the deformation radius or scale of peak mesoscale cascade, is broadly constant across latitudes. The Rhines scale's poor correlation with the arrest scale is unrelated to limited resolution of the dataset at high latitudes. Indeed, both the peak cascade scale (orange lines in Figure~\ref{fig:SUPP:ZonalWithRhines}) and transition scale (black lines in Figure~\ref{fig:SUPP:ZonalWithRhines}) decrease at higher latitudes, as expected.

\subsection*{\MakeUppercase{Maps of Seasonal Energy Transfer}}
\begin{figure}[tbhp]
    \centering
    \includegraphics[scale=1]{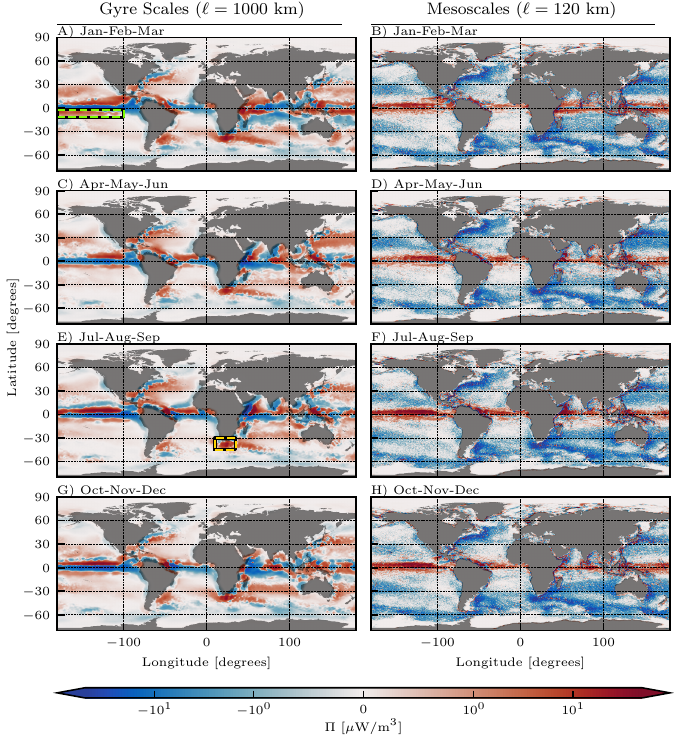}
    \caption{ 
        \textbf{\(\Pi\) Maps at all Seasons}
        Similar to Figure~\ref{fig:SUPP:Pi_maps_1000km}A-D, but showing all four seasons for both gyre-scale and mesoscale transfer:
        [A,B] Jan-Feb-Mar,
        [C,D] Apr-May-Jun,
        [E,F] Jul-Aug-Sep, and
        [G,H] Oct-Nov Dec
        for [A,C,E,G] \(\ell=1000~\)km and [B,D,F,H] \(\ell=120~\)km.
        All panels share a common colour bar, shown along the bottom of the figure.
        All panels show energy scale-transfer arising from the full velocity.
    }
    \label{fig:SUPP:maps:full:seasonal}
\end{figure}

Figure~\ref{fig:SUPP:maps:full:seasonal} presents the analogue of Figure~\ref{fig:SUPP:Pi_maps_1000km}A-D, but showing all four seasons.

\subsection*{\MakeUppercase{Maps of Toroidal Energy Transfer}}
\begin{figure}[tbhp]
    \centering
    \includegraphics[scale=1]{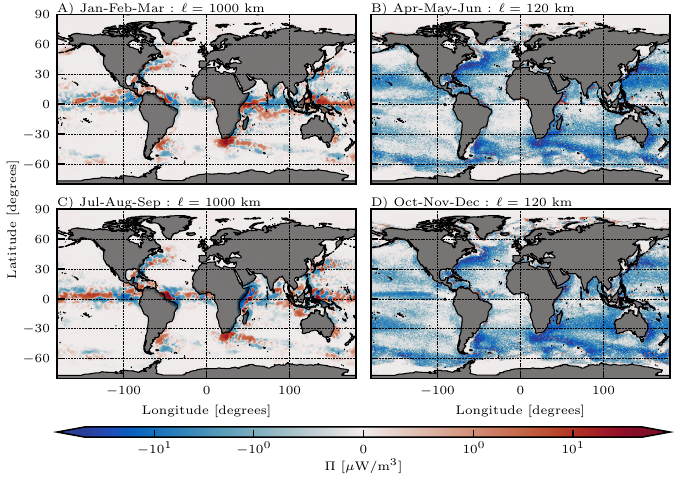}
    \caption{ 
        \textbf{\(\Pi\) Maps for Laterally Non-Divergent Flow}
        Similar to Figure~\ref{fig:SUPP:Pi_maps_1000km}A-D, but showing  the energy scale-transfer arising solely from the laterally non-divergent flow component.
        Panels correspond to 
        [A] Jan-Feb-Mar for \(\ell=1000~\)km, 
        [B] Apr-May-Jun for \(\ell=120~\)km, 
        [C] Jul-Aug-Sep for \(\ell=1000~\)km, 
        [D] Oct-Nov-Dec for \(\ell=120~\)km.
        All panels share a common colour bar, shown along the bottom of the figure.
    }
    \label{fig:SUPP:maps:toroidal}
\end{figure}

Figure~\ref{fig:SUPP:maps:toroidal} presents the analogue of Figure~\ref{fig:SUPP:Pi_maps_1000km}A-D, but for the energy scale-transfer arising solely from the laterally non-divergent flow component.

\subsection*{\MakeUppercase{Comparison with AVISO KE Scale-Tansfer}}

\begin{figure}
    \centering
    \includegraphics[scale=1]{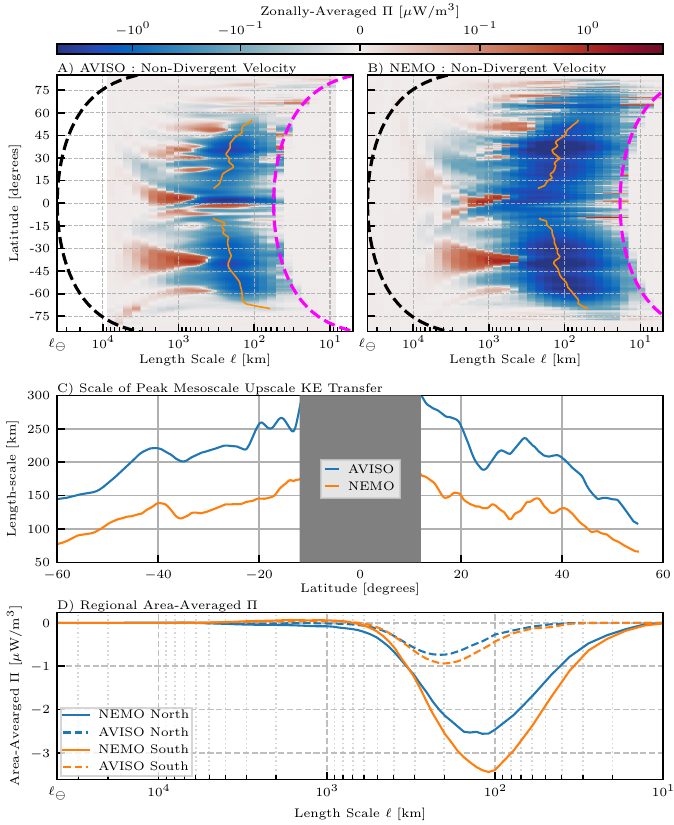}
    \caption{
        \textbf{Comparison of KE Scale-Transfer between NEMO and AVISO}
        Comparison of zonal means of \(\Pi\) between [A] AVISO and [B] NEMO, using the laterally non-divergent (toroidal) velocity components. In [A-B], orange lines show the scale with greatest magnitude, thick dashed black lines show the zonal circumference at each latitude, and thick dashed purple lines show the zonal length of two grid-points at each latitude.
        Panel [C] shows the length-scale of peak mesoscale inverse cascade from each of [A-B].
        Panel [D] shows the area-averaged \(\Pi\) over the north (\([15^\circ\mathrm{N},90^\circ\mathrm{N}]\)) and south (\([90^\circ\mathrm{S},15^\circ\mathrm{S}]\)).
    }
    \label{fig:SUPP:AVISO}
\end{figure}

\begin{figure}
    \centering
    \includegraphics[scale=1.5]{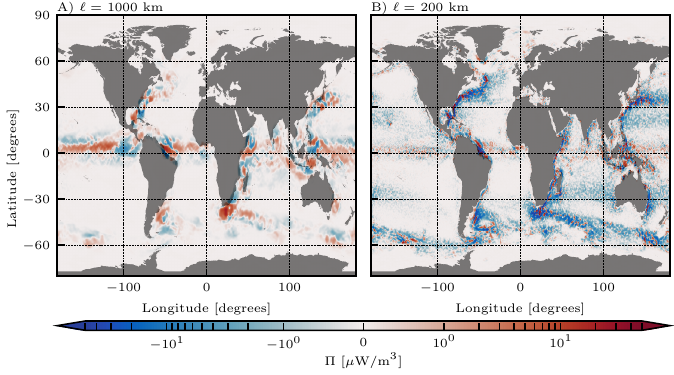}
    \caption{ 
        \textbf{Maps of \(\Pi\) from AVISO} 
        Similar to Figure~\ref{fig:SUPP:Pi_maps_1000km}E-F, but showing the 9-year time-averaged \(\Pi\) obtained from the laterally non-divergent AVISO velocity for [A] 1000~km and [B] 200~km, which roughly corresponds to the peak cascade from Figure~\ref{fig:SUPP:AVISO}D.
        The colour bar is the same as the one used in Figure~\ref{fig:SUPP:Pi_maps_1000km}.
    }
    \label{fig:enter-label}
\end{figure}

Figure~\ref{fig:SUPP:AVISO} below is similar to Fig~\ref{fig:Pi_zonal_means} in the main text and compares the scale-transfer from AVISO to that from NEMO using the laterally non-divergent velocity. Note that velocity field from the AVISO dataset is, by construction, approximately laterally non-divergent \cite{Arbic2014,khatri2018surface} and does not incorporate the Ekman flow component of the oceanic circulation. Panels [A-B] in Figure~\ref{fig:SUPP:AVISO} show similar \(\Pi\) patterns, with even the regions of downscale transfer agreeing well. Panel [C] also shows that in both AVISO and NEMO, the length-scale at which \(\Pi\) peaks generally decreases poleward except in strong current systems.  
    There are two main differences between the NEMO and AVISO datasets: (i) AVISO has a lower \(\Pi\) magnitude and (ii) the peak \(\Pi\) occurs at larger scales (panels C-D). Both of these differences can be attributed to the effective smoothing and lower resolution of the AVISO dataset, as was discussed in prior work\cite{Arbic2013,khatri2018surface}. These differences are also seen in the maps of $\Pi$ using AVISO in Figure~\ref{fig:enter-label}, where the gyre-scale KE transfer shows remarkable agreement with that from NEMO in Figure~\ref{fig:SUPP:Pi_maps_1000km}E, while the mesoscale cascade is weaker compared to that from NEMO in Figure~\ref{fig:SUPP:Pi_maps_1000km}F.    
    Note that unlike KE scale-transfer, comparing KE \emph{spectra} from NEMO and AVISO in \cite{Storer2022} found a remarkably good agreement over all scales $>100~$km.

\end{document}